\documentclass{article}
\usepackage{geometry}
\usepackage[utf8]{inputenc}
\usepackage{cite}
\usepackage{url}
\usepackage{amssymb,amsmath,amsthm}
\usepackage{bm}
\usepackage{graphicx}
\usepackage{enumerate}
\usepackage{microtype}
\usepackage{color}
\usepackage{hyperref}
\usepackage{multirow}
\usepackage{array}
\usepackage{float}
\usepackage{listings}
\usepackage{subcaption}
\usepackage{booktabs}
\usepackage[dvipsnames]{xcolor}
\newcolumntype{C}[1]{>{\centering\let\newline\\\arraybackslash\hspace{0pt}}m{#1}}

\usepackage[ruled,linesnumbered,vlined]{algorithm2e}

\title{Learning-based density-equalizing map}
\author{Yanwen Huang\thanks{Department of Mathematics, The Chinese University of Hong Kong
  ({ywhuang@math.cuhk.edu.hk}).}
\and Lok Ming Lui\thanks{Department of Mathematics, The Chinese University of Hong Kong
  ({lmlui@math.cuhk.edu.hk}).}
\and Gary P. T. Choi\thanks{Department of Mathematics, The Chinese University of Hong Kong
  ({ptchoi@cuhk.edu.hk}).}}

\date{\today}

\begin{document}

\maketitle
\begin{abstract}
Density-equalizing map (DEM) serves as a powerful technique for creating shape deformations with the area changes reflecting an underlying density function. In recent decades, DEM has found widespread applications in fields such as data visualization, geometry processing, and medical imaging. Traditional approaches to DEM primarily rely on iterative numerical solvers for diffusion equations or optimization-based methods that minimize handcrafted energy functionals. However, these conventional techniques often face several challenges: they may suffer from limited accuracy, produce overlapping artifacts in extreme cases, and require substantial algorithmic redesign when extended from 2D to 3D, due to the derivative-dependent nature of their energy formulations. In this work, we propose a novel learning-based density-equalizing mapping framework (LDEM) using deep neural networks. Specifically, we introduce a loss function that enforces density uniformity and geometric regularity, and utilize a hierarchical approach to predict the transformations at both the coarse and dense levels. Our method demonstrates superior density-equalizing and bijectivity properties compared to prior methods for a wide range of simple and complex density distributions, and can be easily applied to surface remeshing with different effects. Also, it generalizes seamlessly from 2D to 3D domains without structural changes to the model architecture or loss formulation. Altogether, our work opens up new possibilities for scalable and robust computation of density-equalizing maps for practical applications. 
\end{abstract}

\section{Introduction}
Density-equalizing maps (DEMs)~\cite{gastner2004diffusion} are spatial transformations that reallocate area in proportion to a given density distribution based on the principle of density diffusion. They were originally developed for cartography, in which it is common to create value-by-area cartograms for mapping and data visualization \cite{sun2010effectiveness,nusrat2016state,hografer2020state}. In particular, under a DEM, different regions in a geographical map are often distorted so that their areas in the new map reflect certain prescribed data values, yet the mapping ideally remains continuous and recognizable. 

DEMs have broad applications across different domains. In cartography, they have been extensively used in recent decades to visualize data such as census results~\cite{gastner2004diffusion}, international factors of democratization~\cite{gleditsch2006diffusion}, housing wealth~\cite{arundel2020divided}, epidemiological spread~\cite{colizza2006role}, climate warming~\cite{lenoir2020species}, trends in information and communication technology~\cite{pratt2012implications}, social media usage~\cite{mislove2011understanding}, and academic publication and citation~\cite{murphy2012neo,pan2012world}. More recently, DEMs have also been found useful for many important tasks in science and engineering. For instance, in medical imaging, density-equalizing map algorithms have been applied to flatten anatomical surfaces or register medical images while preserving area-based quantities of interest~\cite{choi2020area}. In geometry processing, similar principles have been used for surface parameterization and remeshing~\cite{choi2018density}, where a mesh is mapped to a parameter domain with controlled area changes. These examples highlight the utility of DEMs in creating maps or flattenings that faithfully follow the prescribed shape change effects.

Traditionally, DEMs are computed through iterative physics-based procedures. Gastner and Newman~\cite{gastner2004diffusion} introduced an algorithm for generating diffusion-based cartograms: the map is treated as a medium in which a variable (e.g., population density) diffuses over time, causing regions to expand or contract until the density is uniform. This approach solves a series of partial differential equations (PDEs) to incrementally deform the domain. Later, Choi and Rycroft~\cite{choi2018density} extended the density diffusion principle to deform 3D surfaces, flattening them into 2D maps while equalizing a prescribed surface density. Despite the effectiveness of iterative DEM algorithms, they have notable limitations. First, solving the diffusion (or related) equations can be computationally intensive and may require careful tuning for stability and accuracy. High-resolution maps demand many iterations for the density to equalize, and the result might require extra smoothing procedures to avoid extreme local distortions~\cite{gastner2004diffusion}. Moreover, ensuring a bijective (overlap-free) deformation is nontrivial---small mesh overlaps or fold-overs can occur if the deformation is too aggressive or the numerical method is not carefully constrained. Extending these methods to more complex domains is also challenging, typically necessitating custom adaptations of the algorithm~\cite{lyu2024bijective, choi2021volumetric}.

In this paper, we propose a learning-based density-equalizing mapping approach (abbreviated as LDEM) that addresses these challenges. Instead of solving PDEs from scratch for each new dataset, we train a convolutional neural network to learn the mapping that equalizes density. Our method produces high-accuracy density-equalizing mappings for a wide range of prescribed density distributions. Also, the learned model inherently strives to preserve bijectivity, yielding mappings without mesh overlaps. Furthermore, the framework is flexible and can be easily generalized from two dimensions to higher dimensions. Our key contributions are as follows:
\begin{enumerate}
    \item We introduce a novel method for the computation of density-equalizing maps using deep neural networks. In particular, we introduce a loss function involving both the density uniformity and geometric regularity of the mapping. Also, we follow a hierarchical approach to capture both deformations at both the coarse and dense levels.

    \item Our proposed method can handle a wider range of population distributions and achieve a higher mapping accuracy when compared to the traditional diffusion-based approaches, while maintaining low local geometric distortion and bijectivity.

    \item Our method can be easily applied for surface remeshing to achieve different desired remeshing effects.
    
    \item The proposed learning-based framework naturally extends from 2D domains to 3D domains with minimal changes, demonstrating the versatility of the approach.

\end{enumerate}

The rest of this paper is organized as follows. In Section~\ref{sect:previous}, we review previous work on density-equalizing map techniques, including the diffusion-based cartogram and the extension to triangulated surface maps. Section~\ref{sect:background} provides the mathematical background, covering the diffusion equation and the principle of density-equalizing maps, the Beltrami coefficient for quantifying geometric distortion and bijectivity, and an overview of convolutional neural networks. Section~\ref{sect:main} details our proposed learning-based DEM methodology, and Section~\ref{sect:experiment} presents experimental results comparing our method with prior approaches. In Section~\ref{sect:application}, we demonstrate the effectiveness of our method for surface remeshing. In Section~\ref{sect:3d}, we describe the extension of our method for 3D mapping problems. Finally, Section~\ref{sect:conclusion} concludes the paper and discusses future directions.

\section{Previous Work}\label{sect:previous}
Over the past few decades, various computational methods have been developed to construct density-equalizing maps. Below, we highlight two representative approaches and discuss related developments.

\subsection{Diffusion-Based Cartogram}
Conventional methods for generating contiguous cartograms (density-adjusted maps) often required iterative ``rubber-sheet" deformations that could produce overlapping or hard-to-read maps \cite{tobler1973continuous,dougenik1985algorithm,dorling1996area}. In 2004, Gastner and Newman~\cite{gastner2004diffusion} addressed these issues by introducing a diffusion-based cartogram algorithm. In their method, the input density (e.g., population per region) is treated as a fluid that diffuses across the map. As the density flows from higher to lower concentration areas, the map is continuously deformed – regions with surplus density expand while those with deficit contract – in order to conserve the total ``mass" in any region. Formally, the method solves the heat diffusion equation $\partial \rho/\partial t = \nabla^2 \rho$ on the map domain, with appropriate boundary conditions, and concurrently advects the map coordinates using the density flux (ensuring that areas change in proportion to the diffusing mass). The process is run until a steady state is reached where $\rho$ becomes uniform; at that point, the deformation of the map yields a density-equalized cartogram.

This diffusion-based approach produces contiguous, non-overlapping cartograms that largely preserve relative locality and shape, thereby improving readability over earlier methods \cite{gastner2004diffusion}. Gastner and Newman illustrated the technique with examples ranging from electoral maps to disease incidence, showing that it avoided the axis-alignment biases and region overlaps that plagued previous algorithms. One practical consideration is that fine-grained density variations can lead to very local distortions in the map; in practice, a slight smoothing of the input density (e.g., via Gaussian blur) is often applied to maintain cartographic readability at the expense of small accuracy loss \cite{gastner2004diffusion}. The original algorithm also involves computationally intensive PDE integration, typically requiring many time steps on a grid, which can be slow for high-resolution outputs. Subsequent work by Gastner et al. \cite{gastner2018fast} introduced a faster, flow-based implementation that significantly accelerates cartogram generation. Nonetheless, the diffusion-based method set a new standard for contiguous cartograms, and its success spurred further extensions to new domains.

\subsection{Density-Equalizing Maps for Surface Flattening}
Choi and Rycroft~\cite{choi2018density} extended the computation of density-equalizing maps to 3D triangulated surfaces, enabling the flattening of a curved surface into the plane while redistributing area according to a given density function. Their method operates analogously to the planar case: a scalar density defined on the surface is allowed to diffuse over time, and the surface’s parameterization (mapping to 2D) is iteratively updated to reflect the diffusion. In practice, the algorithm starts with an initial flattening of the surface (for example, a conformal map of the 3D surface onto a 2D domain) and defines a density $\rho$ on the surface (such as one proportional to some scalar field or the area element). As $\rho$ diffuses across the surface, the 2D coordinates of the mesh vertices are adjusted so that areas in the parameterization change in proportion to the mass flow. By the end of the process, the surface is flattened in such a way that regions with higher initial density occupy larger areas in the planar map. This allows, for instance, one to obtain an area-preserving parameterization by simply setting $\rho$ based on the area element of the surface. Choi and Rycroft demonstrated applications of the surface DEM method in data visualization and surface remeshing, laying the groundwork for subsequent improvements in surface and volumetric domains.

While effective, the surface DEM algorithm shares some limitations with its planar predecessor. It requires solving a time-dependent PDE on a triangle mesh, which can be computationally heavy for complex or high-resolution surfaces. The method is inherently restricted to simply connected open surfaces and does not directly handle domains with other topologies. Additionally, large variations in the prescribed density can cause significant distortion of the mesh, and ensuring that the flattening remains bijective (without any fold-overs) is nontrivial. In practice, careful implementation and possibly mesh refinement are needed to minimize element inversion. 

\subsection{Other Related Works}
Beyond the above approaches, there have been numerous refinements and extensions of DEM techniques. In cartography, researchers have also examined the usability and perception of cartograms; for instance, user studies have evaluated the effectiveness of contiguous area cartograms in conveying information~\cite{arranz2021eval}. On the algorithmic side, Gastner et al.’s flow-based method \cite{gastner2018fast} significantly improved the runtime of diffusion cartograms, making high-resolution density-equalizing maps more practical. DEM algorithms have also been adapted to different domain topologies. Li and Aryana~\cite{li2018diffusion} proposed a diffusion-based density-equalizing map for spherical surfaces, effectively creating cartograms on the globe. More recently, the computation of surface density-equalizing maps has also been developed for genus-0 closed surfaces~\cite{lyu2024spherical,lyu2024ellipsoidal} and surfaces with other topologies~\cite{lyu2024bijective,yao2025toroidal,shaqfa2025disk,choi2025hemispheroidal}. In addition, volumetric extensions have been explored: Li and Aryana~\cite{li2019visualization} developed a volumetric method for handling subsurface data. Choi and Rycroft~\cite{choi2021volumetric} also presented a general density-equalizing mapping framework for volumetric datasets, deforming a 3D volume so that a given density becomes uniform throughout the interior. Each of these extensions addresses specific challenges (such as enforcing bijectivity or handling complex geometries) with tailored algorithmic modifications. However, all of the above methods still rely on solving physical diffusion equations or iterative optimizations for each new input, underscoring the need for more efficient solutions. This has motivated interest in data-driven approaches, as we propose in this paper, to learn the mapping function directly, bypassing the costly iterative solves for each instance.

\section{Mathematical Background} \label{sect:background}

In this section, we review the key mathematical tools underpinning our proposed learning‐based method for density-equalizing maps. These include the basic concepts of the diffusion equation and density-equalizing maps, the Beltrami coefficient in quasi-conformal theory, and convolutional neural networks.

\subsection{Diffusion Equation and Density-Equalizing Maps}
The diffusion equation models how a density distribution $\rho(\mathbf{x},t)$ spreads over time:
\begin{equation}
\frac{\partial \rho}{\partial t}(\mathbf{x},t)
= \Delta \rho(\mathbf{x},t),
\quad \mathbf{x}\in \Omega\,,
\end{equation}
where $\Delta$ is the Laplacian operator. We impose the no-flux boundary conditions
\begin{equation}
\nabla \rho \cdot \mathbf{n} = 0
\quad\text{on }\partial\Omega,
\end{equation}
where $\mathbf{n}$ is the unit outward normal, so that mass is conserved within the domain. 

In density-equalizing maps, given a density distribution $\rho$, the above diffusion equation is solved iteratively, with the density gradient driving the shape deformation of the domain:
\begin{equation}
    \mathbf{u} = - \frac{\nabla \rho}{\rho},
\end{equation}
so that the displacement of the vertices in the domain satisfies
\begin{equation}
    \mathbf{x}(t) = \mathbf{x}(0) + \int_0^t \mathbf{u}(\mathbf{x}(\tau),\tau) d \tau.
\end{equation}
As $t\to\infty$, the solution converges to
\begin{equation}
\rho(\mathbf{x},t)\;\longrightarrow\;
\bar{\rho}
=\frac{1}{|\Omega|}\int_\Omega \rho(\mathbf{x},0)\,d\mathbf{x},
\end{equation}
i.e.\ a uniform density. In other words, this uniformization drives the domain deformation so that different regions expand or contract based on $\rho$, yielding a transformation with prescribed area changes.

\subsection{Beltrami Coefficient}
A quasi-conformal map $f:\Omega \subset \mathbb{C} \to \mathbb{C}$ satisfies the Beltrami equation
\begin{equation}
    f_{\bar z} = \mu(z) f_{z},
\end{equation}
where $f_z$ and $f_{\bar z}$ are the Wirtinger derivatives
\begin{equation}
f_z = \tfrac12\bigl(f_x - i\,f_y\bigr),
\quad
f_{\bar z} = \tfrac12\bigl(f_x + i\,f_y\bigr),
\end{equation}
and $\mu(z)$ is a complex-valued function (called the Beltrami coefficient) with $\|\mu\|_{\infty} < 1$. In other words, we have
\begin{equation}
\mu(z) = \frac{f_{\bar z}}{f_z}.
\end{equation}
Intuitively, a quasi-conformal map $f$ maps infinitesimal circles to infinitesimal ellipses, with the local aspect ratio change expressed as
\begin{equation}
K(f) = \frac{1 + |\mu|}{1 - |\mu|}.
\end{equation}
Therefore, $|\mu|$ can effectively measure the quasi-conformal distortion of a mapping, with $|\mu|=0$ if and only if $f$ is conformal. Also, any local fold-overs can be captured by a value of $|\mu|>1$. In other words, $|\mu|$ also provides us with a simple way for assessing the bijectivity of the mapping results produced by our method.

\subsection{Convolutional Neural Network (CNN)}
In our proposed method, the convolutional neural network (CNN) is employed to approximate the density transformation. CNNs are a class of deep neural networks designed to process data with grid-like structures, such as images or volumetric data. A CNN applies a series of convolution operations, which compute the weighted sum of local neighborhoods, to extract hierarchical features from the input data~\cite{Goodfellow2016deep}.

Mathematically, a convolution operation between an input $I(x)$ and a kernel $K(x)$ is defined as:
\begin{equation}
(F * K)(x) = \int_{\mathbb{R}^d} I(u)K(x-u) \, du,
\end{equation}
where $F * K$ denotes the convolution, $d$ is the dimensionality of the input space, and $K(x)$ represents the learnable filter or kernel. For discrete data, this becomes:
\begin{equation}
(F * K)[i] = \sum_{j} I[j]K[i-j],
\end{equation}
where the summation is over the neighborhood defined by the kernel size.

The network is composed of multiple layers, including convolutional layers, activation functions, pooling layers, and fully connected layers. Specifically, the \emph{Convolutional Layers} apply filters to extract spatial features. The \emph{Activation Functions} introduce non-linearity, commonly using ReLU:
\begin{equation}
\sigma(x) = \max(0, x).
\end{equation}
The \emph{Pooling Layers} reduce the spatial dimension for computational efficiency, typically using max pooling:
\begin{equation}
P(x) = \max_{i \in \mathcal{N}(x)} F[i],
\end{equation}
where $\mathcal{N}(x)$ is the neighborhood of $x$.
The \emph{Fully Connected Layers} combine features for final predictions.

The CNN is trained to minimize a loss function, typically defined as the mean squared error (MSE) between the predicted and target values:
\begin{equation}
\mathcal{L} = \frac{1}{N} \sum_{i=1}^N \left( v_{\text{predicted}}(x_i) - v_{\text{target}}(x_i) \right)^2,
\end{equation}
where $N$ is the number of training samples, and $v_{\text{predicted}}(x_i)$ and $v_{\text{target}}(x_i)$ are the predicted and target values at $x_i$, respectively.

By leveraging the ability of CNNs to learn complex transformations, our method effectively computes density-equalizing maps with improved scalability and generalizability.

\section{Proposed Method} \label{sect:main}

\begin{figure}[t!]
    \centering
    \includegraphics[width=0.8\textwidth]{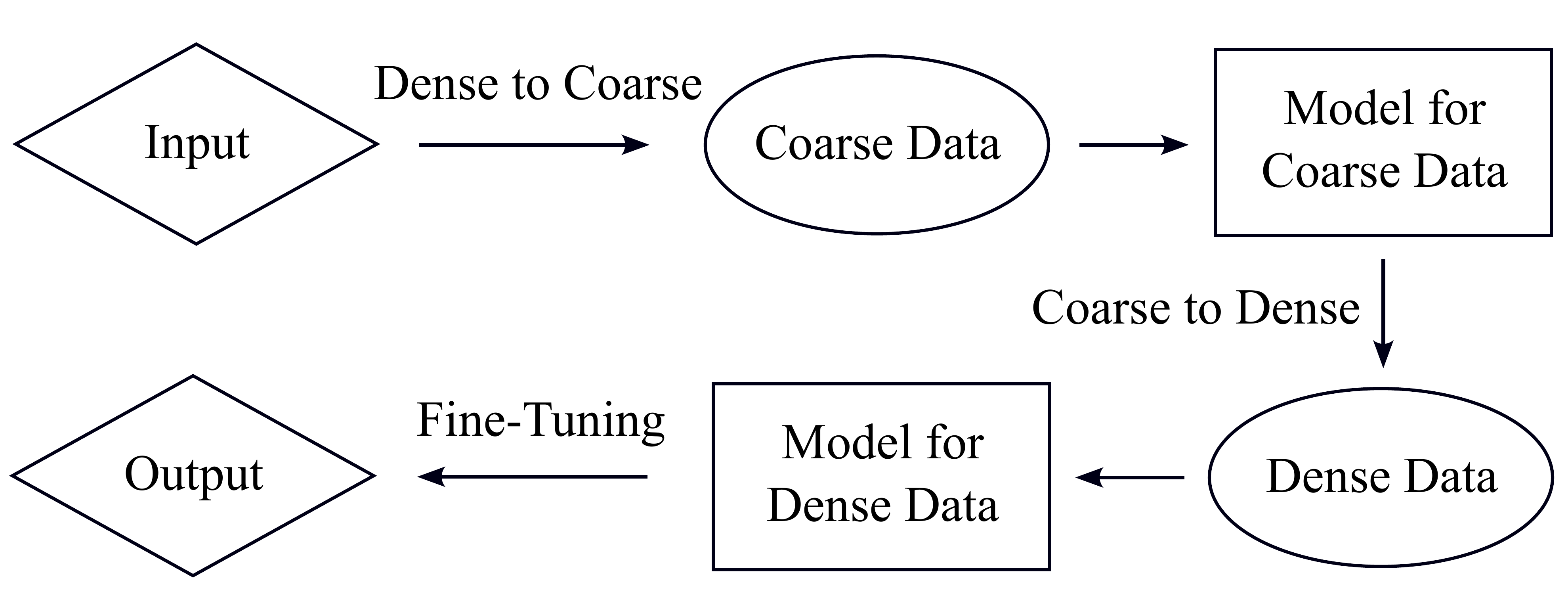}
    \caption{An overview of the proposed learning-based density-equalizing mapping (LDEM) method.}
    \label{fig:flowchart}
\end{figure}

The proposed learning-based density-equalizing mapping method (LDEM) follows a hierarchical pipeline consisting of five main stages, as illustrated in Fig.~\ref{fig:flowchart}:

\begin{enumerate}
    \item \textbf{Data Initialization}: The process begins with an input map represented on an $n \times n$ meshgrid. Population values are assigned to each triangular element formed by a triangulation of this grid.
    \item \textbf{Dense-to-Coarse Transformation}: The fine-grained input is downsampled into a coarser representation to reduce computational complexity and capture global structural information.
    \item \textbf{Coarse Model Processing}: A neural network model processes the coarse data to produce a preliminary transformation field.
    \item \textbf{Coarse-to-Dense Transformation}: The intermediate output is upsampled using an interpolation scheme, transferring the coarse prediction back to the original resolution.
    \item \textbf{Fine-Tuning with Dense Model}: A separate dense-level model further refines the interpolated output, resulting in a final transformation map that achieves high spatial accuracy.
\end{enumerate}
In the following sections, we will describe the five stages in detail.

\subsection{Data Initialization}

\begin{figure}[t!]
    \centering
    \begin{subfigure}[t]{0.48\linewidth}
        \centering
        \includegraphics[width=\linewidth]{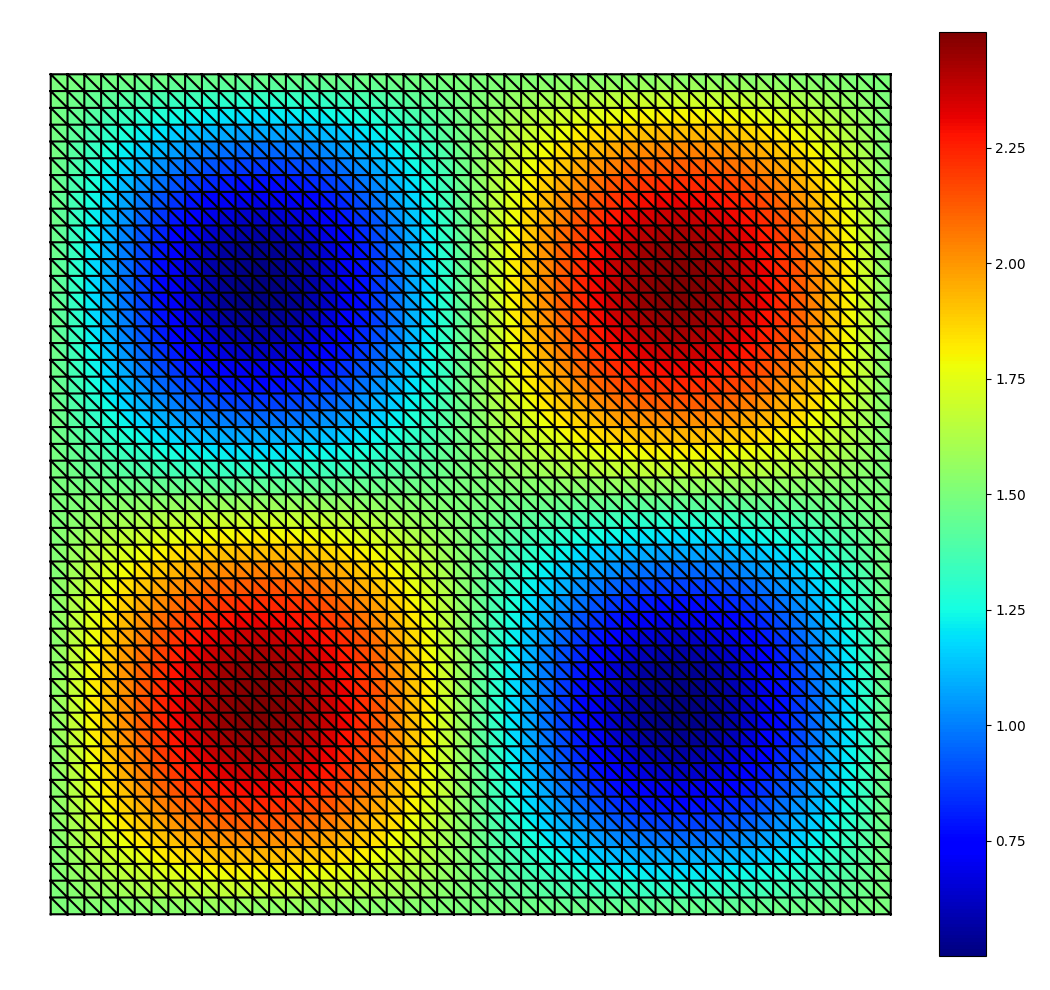}
        \caption{Initial domain}
        \label{fig:initial_map}
    \end{subfigure}
    \hfill
    \begin{subfigure}[t]{0.48\linewidth}
        \centering
        \includegraphics[width=\linewidth]{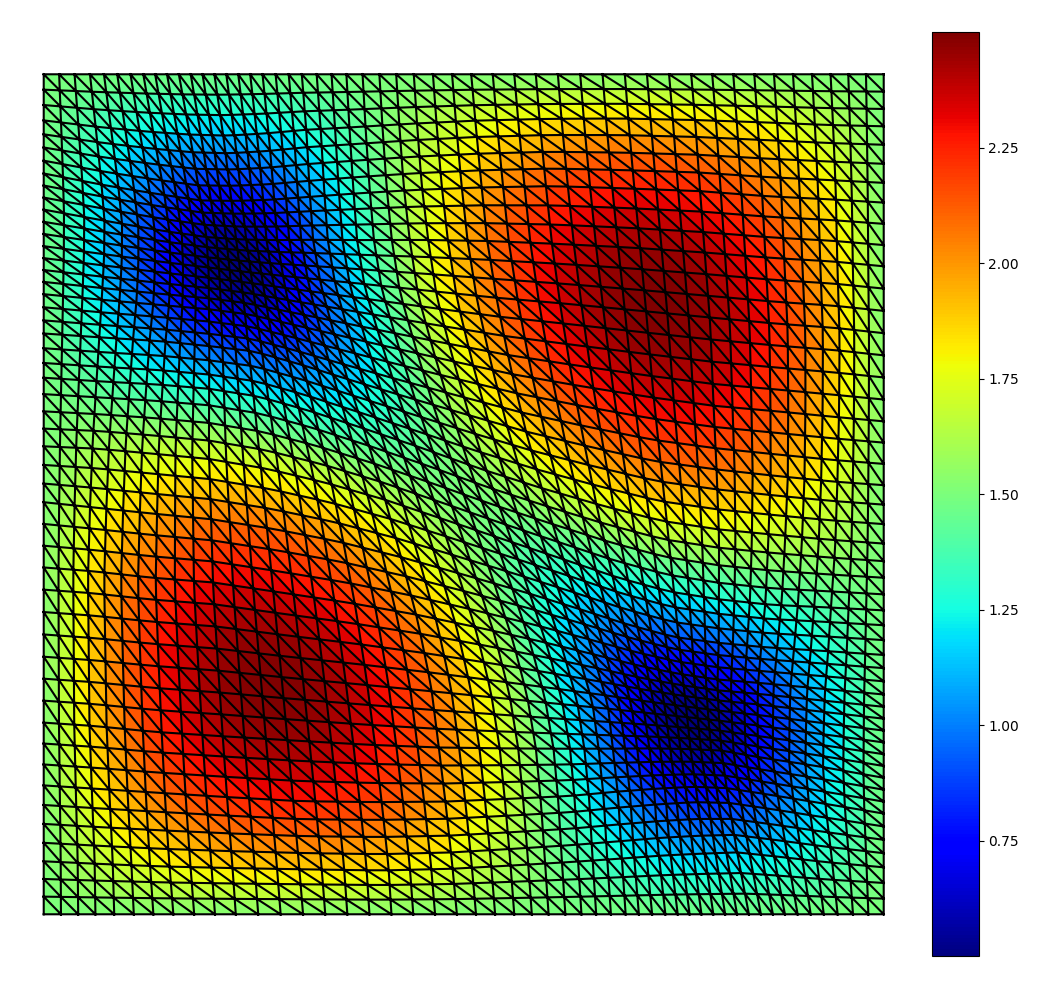}
        \caption{Final density-equalizing mapping result}
        \label{fig:transformed_map}
    \end{subfigure}
    \caption{An illustration of density-equalizing maps. Both the initial domain and final mapping result are triangulated and color-coded with the given population. The transformation of the initial triangulated domain adjusts the vertex positions so that the population density, i.e., the population per unit area in the deformed domain, is equalized.}
    \label{fig:initial_and_transformed_maps}
\end{figure}

We initialize the domain using a regular $n \times n$ meshgrid. A triangulation is applied to divide the domain into non-overlapping triangles, each associated with a given population (a real positive scalar). An example of the initial configuration is shown in Fig.~\ref{fig:initial_map}.

The objective of the transformation is to relocate the mesh points such that the resulting map equalizes the population density: each triangle is transformed in a way that its area becomes approximately proportional to the original population weight, effectively producing a near-uniform density across the domain. Fig.~\ref{fig:transformed_map} illustrates the result of this transformation. Visually, the adjusted triangles vary in size to accommodate different population levels, while maintaining continuity and avoiding overlaps.

\subsection{Dense-to-Coarse Transformation}

To improve computational efficiency, we first approximate the dense input data using a coarser representation. This step enables fast processing while retaining essential global structures of the original data distribution. 

We start by generating a coarse grid in the domain. Specifically, a coarse grid of resolution \( D_{\text{coarse}} \times D_{\text{coarse}} \) is defined over the unit square \([0,1]^2\), with uniformly spaced vertices:
\begin{equation}\label{eqt:coarsegrid}
x_i, y_j \in \left\{ 0, \frac{1}{D_{\text{coarse}}-1}, \ldots, 1 \right\}, \quad i, j = 1, 2, \ldots, D_{\text{coarse}}.
\end{equation}
The vertex set is denoted as:
\begin{equation}
\mathbf{v}_{\text{coarse}} = \left\{ (x_i, y_j) \left| \right. \  i, j = 1, 2, \ldots, D_{\text{coarse}} \right\}.
\end{equation}

Then, each square cell in the coarse grid is divided into two right-angled triangles. Let \( k = i \cdot D_{\text{coarse}} + j \) denote the linear index of the grid point \((i,j)\). Then the two triangles covering the cell are:
\begin{equation}
\begin{aligned}
\mathbf{f}_{k,1} &= [k, k+1, k+D_{\text{coarse}}], \\
\mathbf{f}_{k,2} &= [k+1, k+D_{\text{coarse}}+1, k+D_{\text{coarse}}].
\end{aligned}
\end{equation}
After performing the triangulation of the coarse grid, for each triangle \(\mathbf{f}_i = [\mathbf{v}_a, \mathbf{v}_b, \mathbf{v}_c]\), we calculate the centroid by:
\begin{equation}
\mathbf{c}_i = \frac{\mathbf{v}_a + \mathbf{v}_b + \mathbf{v}_c}{3}.
\end{equation}

Next, we perform the following population aggregation step to define the population on the coarse grid. Given dense triangle centroids \(\mathbf{c}_{\text{dense}}\) with corresponding population values \(\mathbf{p}_{\text{dense}}\), the population value for each coarse triangle is computed via neighborhood averaging. For a coarse centroid \(\mathbf{c}_{\text{coarse},i}\), we define its neighborhood as:
\begin{equation}
\mathcal{N}_i = \left\{ j \mid \left\| \mathbf{c}_{\text{dense},j} - \mathbf{c}_{\text{coarse},i} \right\| < \frac{1}{D_{\text{coarse}}} \right\}.
\end{equation}
The aggregated population is then:
\begin{equation}
\mathbf{p}_{\text{coarse},i} = \frac{1}{|\mathcal{N}_i|} \sum_{j \in \mathcal{N}_i} \mathbf{p}_{\text{dense},j}.
\end{equation}
This preserves local density structure while significantly reducing resolution.

\subsection{Coarse Model Processing} \label{sect:coarsemodel}

To learn the transformation fields on the coarse and dense grids, we design an efficient neural network architecture with bottleneck and sparse convolutional layers. The model is lightweight, flexible, and effective for grid-based population data. 

\subsubsection{Model Configuration}
Below, we describe the configuration of our model. 

As for the input and output dimensions, the model receives a vector of population values as input and outputs a corresponding transformation vector. Input/output dimensions are adjusted according to the grid resolution.

For the dimensionality reduction, a fully connected layer reduces the input dimension \(I\) to a low-dimensional bottleneck space \(B\), typically set as \(B=1\):
    \begin{equation}
    \mathbf{z} = \sigma(\mathbf{W}_1 \mathbf{x} + \mathbf{b}_1), \quad \mathbf{z} \in \mathbb{R}^{B}.
    \end{equation}
    Here, \(\sigma\) denotes the sigmoid activation.

The model performs sparse processing via 1D convolution. Specifically, a 1D convolution layer with group-wise operations is used to extract nonlinear features:
\begin{equation}
\mathbf{h} = \text{ReLU}(\text{Conv1D}(\mathbf{z})).
\end{equation}
The feature vector is then mapped back to the original output dimension using another fully connected layer:
\begin{equation}
\mathbf{y} = \mathbf{W}_2 \mathbf{h} + \mathbf{b}_2.
\end{equation}
The end-to-end computation of the model is summarized as:
\begin{equation}
\mathbf{y} = \mathbf{W}_2 \cdot \text{ReLU}\left( \text{Conv1D}\left( \sigma(\mathbf{W}_1 \mathbf{x} + \mathbf{b}_1) \right) \right) + \mathbf{b}_2.
\end{equation}
Two independent models are constructed for coarse and dense grid processing. Here we first focus on the problem on the coarse grids. Later, we will extend it for the dense model.

Altogether, this configuration ensures the model is both scalable and adaptable to grid resolution changes, supporting the hierarchical structure of the proposed framework.

\subsubsection{Training Process} 
\label{lossdescription}

The model training process consists of three key components: loss function design, model initialization, and multi-stage training.

As for the loss function, recall that density-equalizing maps aim to create shape deformations based on the prescribed population. Therefore, one fundamental element in the loss function is the density equalization measure. Also, it is desired that the shape deformations are smooth and natural, without sharp changes or even mesh fold-overs. Hence, our proposed training objective combines population consistency and geometric regularity criteria. 

Specifically, consider a $N \times N$ coarse grid in a 2D domain discretized into $2(N-1)^2$ triangles, in which a population value $p_i$ is defined on each triangle $t_i$. The density equalization aims to produce a new configuration where the area of each transformed triangle $t'_i$ is proportional to the associated population $p_i$, thereby yielding a uniform density. To assess the uniformity of the density, we define the \textit{Density Uniformity Loss} as follows:
\begin{equation}
    \mathcal{L}_{\text{density}} = \frac{\mathrm{std}(\boldsymbol{\rho}_f)}{\mathrm{mean}(\boldsymbol{\rho}_f)}, 
\end{equation}
where $\boldsymbol{\rho}_f = (\rho_{f,1}, \rho_{f,2} , \dots, \rho_{f,2(N-1)^2})$ is the vector of face-wise densities defined by
\begin{equation}
     \rho_{f,i} = \frac{p_i}{\text{Area}(t'_i)},
\end{equation}
with $i = 1, 2, \dots, 2(N-1)^2$. It is easy to see that $\mathcal{L}_{\text{density}} = 0$ if and only if the density is fully equalized in the entire domain.

Next, to ensure that the shape deformation is smooth and natural, we consider the following two \textit{Geometric Regularization Loss} terms:
\begin{equation}\label{eqt:slope}
    \mathcal{L}_{\text{slope}} = \frac{1}{N} \sum_{i=1}^{N} \left( \text{slope}_{x,i} + \text{slope}_{y,i} \right)
\end{equation}
and
\begin{equation}\label{eqt:distance}
    \mathcal{L}_{\text{distance}} = \frac{1}{N} \sum_{i=1}^{N} \left( \text{distance}_{x,i} + \text{distance}_{y,i} \right),
\end{equation}
where the terms are defined as follows:
\begin{itemize}
    \item $\text{slope}_{x,i} = \sum_{j=1}^{N-2} \left| s_{j+1}^{(x)} - s_j^{(x)} \right|$, where $s_j^{(x)} = \dfrac{y_{j+1} - y_j}{x_{j+1} - x_j + \varepsilon}$ is the slope between consecutive points in the $i$-th group along the $x$-axis. Here, $\varepsilon = 10^{-8}$ is a small constant to ensure numerical stability in slope calculations.
    
    \item $\text{slope}_{y,i} = \sum_{j=1}^{N-2} \left| s_{j+1}^{(y)} - s_j^{(y)} \right|$, where $s_j^{(y)} = \dfrac{x_{j+1} - x_j}{y_{j+1} - y_j + \varepsilon}$ is the slope between interleaved points in the $i$-th group along the $y$-axis, and $\varepsilon = 10^{-8}$.

    \item $\text{distance}_{x,i} = \sum_{j=1}^{N-2} \left| d_{j+1}^{(x)} - d_j^{(x)} \right|$, where $d_j^{(x)} = (x_{j+1} - x_j)^2 + (y_{j+1} - y_j)^2$ is the squared distance between neighboring points in $x$-axis groups.

    \item $\text{distance}_{y,i} = \sum_{j=1}^{N-2} \left| d_{j+1}^{(y)} - d_j^{(y)} \right|$, where $d_j^{(y)} = (x_{j+1} - x_j)^2 + (y_{j+1} - y_j)^2$ is the squared distance between neighboring points in $y$-axis groups (arranged interleaved).
\end{itemize}
Intuitively, the terms $\text{slope}_{x,i}$ and $\text{slope}_{y,i}$ in $\mathcal{L}_{\text{slope}}$ effectively capture the spatial variation of the slope of the edges in each cell in the grid. Also, the terms $\text{distance}_{x,i}$ and $\text{distance}_{y,i}$ in $\mathcal{L}_{\text{distance}}$ assess the spatial variation in the length of the edges in each cell in the grid. By minimizing $\mathcal{L}_{\text{slope}}$ and $\mathcal{L}_{\text{distance}}$, we can ensure a smooth change across the entire domain, thereby effectively enhancing the geometric regularity and naturally avoiding mesh fold-overs.

Combining the above loss functions, we have the overall loss function for training as follows:
\begin{equation} \label{eqt:custom}
\mathcal{L} = \lambda_d \cdot \mathcal{L}_{\text{density}} + \lambda_s \cdot \mathcal{L}_{\text{slope}} + \lambda_l \cdot \mathcal{L}_{\text{distance}},
\end{equation}
where $\lambda_d, \lambda_s , \lambda_l$ are nonnegative parameters.

As for the model initialization, in the initialization phase, we use the Adam optimizer and the mean squared error (MSE) loss to establish a stable baseline.

Finally, as for the model training phases, the model is first trained with MSE loss in the initialization phase:
\begin{equation}
\mathcal{L}_{\text{init}}^{\text{coarse}}  = \frac{1}{D_{\text{coarse}}} \sum_{i=1}^{D_{\text{coarse}}} (\hat{\mathbf{y}}_i - \mathbf{y}_i)^2.
\end{equation}
Here, \( \mathbf{y}_i \in \mathbb{R}^2 \) denotes the target output coordinates corresponding to the identity mapping, and \( \hat{\mathbf{y}}_i \) is the model's predicted output for the input point \( \mathbf{x}_i \). In this initialization process, the goal is to encourage the network to behave approximately as the identity function over the spatial domain, i.e., \( \hat{\mathbf{y}}_i \approx \mathbf{x}_i \). This provides a stable starting point for subsequent training phases that enforce density equalization, helping to prevent large initial distortions in the learned transformation. Then, in the fine-tuning phase, training proceeds using the proposed loss function $\mathcal{L}$ in Eq.~\eqref{eqt:custom}. Gradient clipping is applied in both stages to enhance training stability.

Altogether, the above coarse model produces a preliminary coarse transformation field, which is then used in the subsequent stages.

\subsection{Coarse-to-Dense Transformation}

To bridge the coarse and dense resolutions, we employ interpolation techniques~\cite{weiser1988note} that transfer the transformation field predicted on the coarse grid to the dense grid. This enables high-resolution refinement while preserving coarse-level consistency.

First, the coarse function values \( f_{\text{coarse}}(x_i, y_j) \) are defined on the coarse grid in Eq.~\eqref{eqt:coarsegrid}. Then, we construct a bilinear interpolation function:
\begin{equation}
f(x, y) = \sum_{i,j} w_{ij}(x, y) \cdot f_{\text{coarse}}(x_i, y_j),
\end{equation}
where \( w_{ij}(x, y) \) are bilinear weights based on the proximity of \((x, y)\) to coarse grid nodes. A dense grid is defined as:
\begin{equation}
x_k, y_l \in \left\{ 0, \frac{1}{D_{\text{dense}}-1}, \ldots, 1 \right\}, \quad k, l = 1, 2, \ldots, D_{\text{dense}}.
\end{equation}
The interpolated function values on this grid are:
\begin{equation}
f_{\text{dense}}(x_k, y_l) = f(x_k, y_l).
\end{equation}
Finally, the interpolation is applied to both coordinate components independently:
\begin{equation}
\begin{split}
X_{\text{dense}}(x_k, y_l) &= f_x(x_k, y_l), \\
Y_{\text{dense}}(x_k, y_l) &= f_y(x_k, y_l),
\end{split}
\end{equation}
where
\begin{equation}
\begin{split}
f_x(x,y) &= \sum_{i,j} w_{ij}(x,y) \cdot [f_{\text{coarse}}(x_i, y_j)]_1, \\
f_y(x,y) &= \sum_{i,j} w_{ij}(x,y) \cdot [f_{\text{coarse}}(x_i, y_j)]_2,
\end{split}
\end{equation}
with $[\cdot]_1$ and $[\cdot]_2$ denoting the first and second components respectively. The final dense transformation grid is then:
\begin{equation}
\mathbf{V}_{\text{dense}} = \left\{ \left(X_{\text{dense}}(x_k, y_l), Y_{\text{dense}}(x_k, y_l)\right) | \ k, l = 1, 2, \ldots, D_{\text{dense}} \right\}.
\end{equation}

\subsection{Fine-Tuning with Dense Model}
After the interpolation, we use a separate dense-level model to fine-tune the interpolated output and result in the final transformation. The model undergoes a two-stage training process similar to that of the coarse model in Section~\ref{sect:coarsemodel}.

First, an initialization phase is conducted using the MSE loss to bring the dense model close to the interpolated map. The objective is to preserve spatial coherence and ensure a smooth starting point for subsequent fine-tuning. This phase uses a relatively high learning rate to allow faster convergence:
\begin{equation}
\mathcal{L}_{\text{init}}^{\text{dense}} = \frac{1}{D_{\text{dense}}} \sum_{i=1}^{D_{\text{dense}}} (\hat{\mathbf{Y}}_i - \mathbf{Y}_i)^2,
\end{equation}
where $\mathbf{Y}_i \in \mathbb{R}^2$ denotes the interpolated coordinates from the coarse model, and \( \hat{\mathbf{Y}}_i \) is the predicted output from the dense model.

Following this, we perform a fine-tuning phase with a smaller learning rate to refine the transformation. The training objective again uses the same customized loss function defined in Section~\ref{lossdescription}, which balances density-equalization accuracy and geometric regularity. The fine-tuning stage allows the model to capture finer spatial details and reduce residual distortion in the transformation output.

This two-phase training procedure enables the dense model to achieve high-fidelity density-equalizing transformations while remaining consistent with the initialization provided by the coarse model.

\subsection{Hyperparameter Settings}
To ensure stable and effective learning, we configure hyperparameters specifically for the coarse and dense models. Each model undergoes two training phases: an initialization phase and a main training phase. 

For the coarse model, we have the following settings:
\begin{itemize}
    \item For the initialization phase, we set the learning rate as \( \text{init\_lr\_coarse} = 1 \times 10^{-2} \) and the number of epochs as \( \text{init\_epochs\_coarse} = 800 \). This stage provides a well-conditioned starting point for later optimization.
    
    \item For the training phase, we set the learning rate as \( \text{train\_lr\_coarse} = 3 \times 10^{-3} \), the maximum number of epochs as \( \text{max\_epochs\_coarse} = 5000 \), the early stopping patience as 500 epochs, the minimum improvement threshold as \( \text{min\_delta} = 1 \times 10^{-4} \), and the Warm-up period as 150 epochs (during which early stopping is disabled). Early stopping is used to prevent overfitting, and the warm-up period helps stabilize training before convergence is monitored.
\end{itemize}

For the dense model, we have the following settings:
\begin{itemize}
    \item For the initialization phase, we set the learning rate as \( \text{init\_lr\_dense} = 1 \times 10^{-2} \) and the number of epochs as \( \text{init\_epochs\_dense} = 800 \).

    \item For the training phase, we set the learning rate as \( \text{train\_lr\_dense} = 2 \times 10^{-4} \) and the number of epochs as \( \text{train\_epochs\_dense} = 300 \). Here, the lower learning rate ensures fine-grained adjustments to high-resolution predictions.
\end{itemize}

These hyperparameters are empirically selected to balance convergence speed, stability, and final model accuracy for both coarse and dense transformations.

\section{Experimental results} \label{sect:experiment}
\subsection{Experimental Setup}
The proposed method is implemented in Python, leveraging the PyTorch deep learning framework for model definition, training, and optimization. Key packages used in our implementation include \texttt{torch}, \texttt{torch.nn}, \texttt{torch.optim} for neural network components and training routines, \texttt{numpy} for numerical array operations and grid generation, and \texttt{matplotlib.pyplot} for visualization.

In the following experiments, we generate test cases by creating coarse and dense grids and defining different population distributions on their triangular faces. The grid dimensions are set as \(D_{\text{dense}} = 51\) for the dense grid and \(D_{\text{coarse}} = 16\) for the coarse grid. For the hyperparameters in the loss function, we choose \(\lambda_d = D_{\text{coarse}}\), \(\lambda_s = 1\), and \(\lambda_l = 10\) for the coarse model, and \(\lambda_d = D_{\text{dense}}\), \(\lambda_s = 1\), and \(\lambda_l = 10\) for the dense model.

\subsection{Test cases and results}
\begin{figure}[t!]
    \centering
    \begin{subfigure}[b]{0.31\textwidth}
        \includegraphics[width=\textwidth]{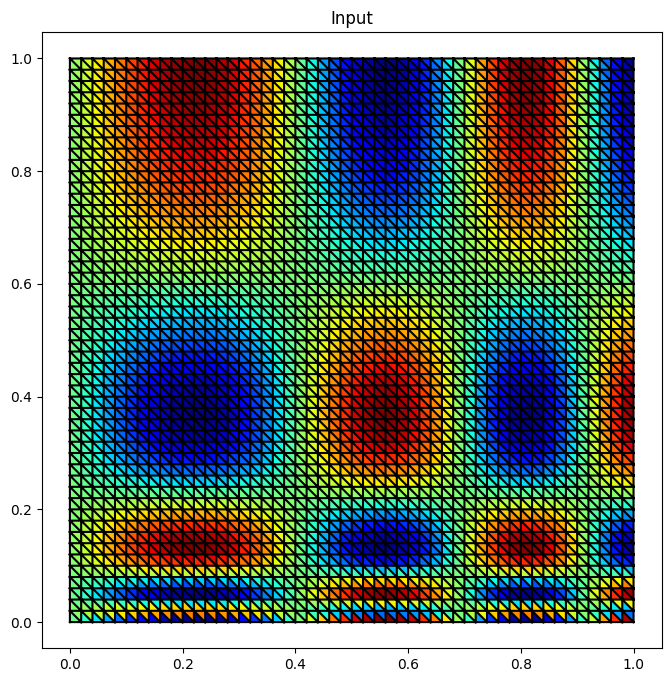}
        \caption{Input}
    \end{subfigure}
    \begin{subfigure}[b]{0.31\textwidth}
        \includegraphics[width=\textwidth]{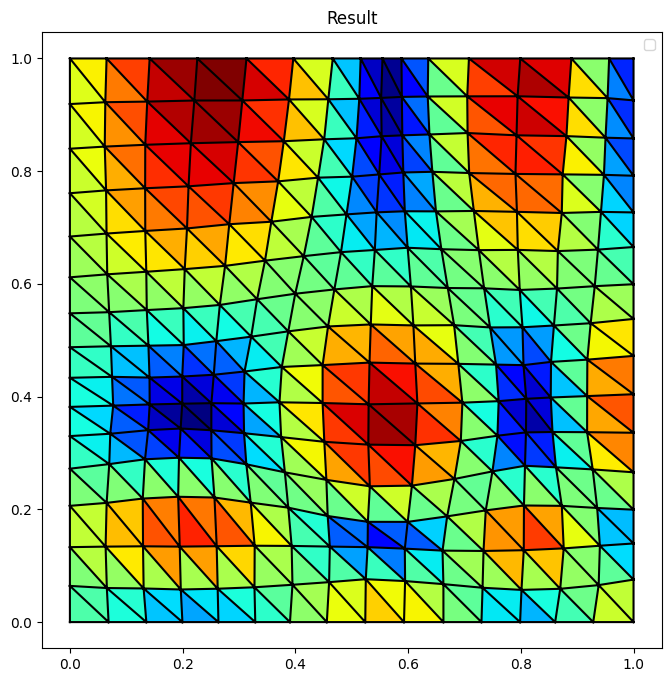}
        \caption{Coarse Model Output}
    \end{subfigure}
    \begin{subfigure}[b]{0.35\textwidth}
        \includegraphics[width=\textwidth]{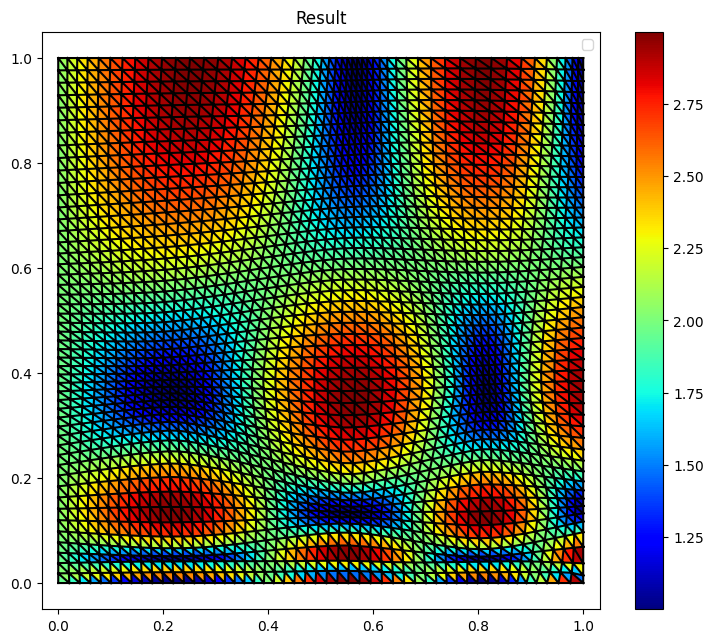}
        \caption{Dense Model Output}
    \end{subfigure}
    \caption{The experimental result obtained by our proposed LDEM method for the \emph{Complex Sinusoidal Variation} test case.}
    \label{fig:complex_sinusoidal_results}
\end{figure}

\begin{figure}[t!]
    \centering
    \begin{subfigure}[b]{0.31\textwidth}
        \includegraphics[width=\textwidth]{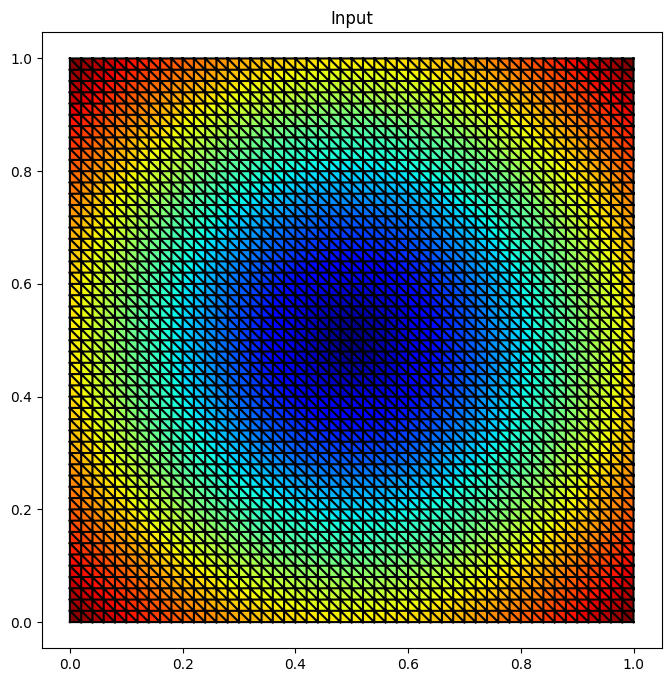}
        \caption{Input}
    \end{subfigure}
    \begin{subfigure}[b]{0.31\textwidth}
        \includegraphics[width=\textwidth]{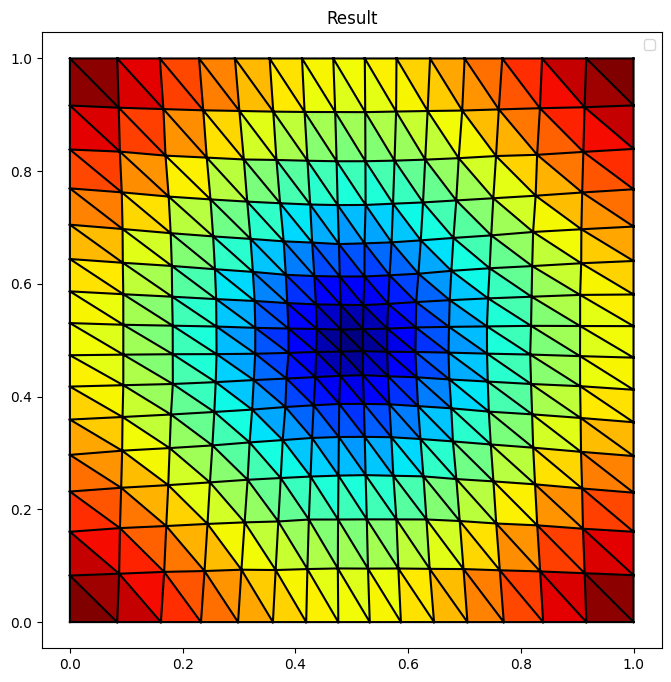}
        \caption{Coarse Model Output}
    \end{subfigure}
    \begin{subfigure}[b]{0.35\textwidth}
        \includegraphics[width=\textwidth]{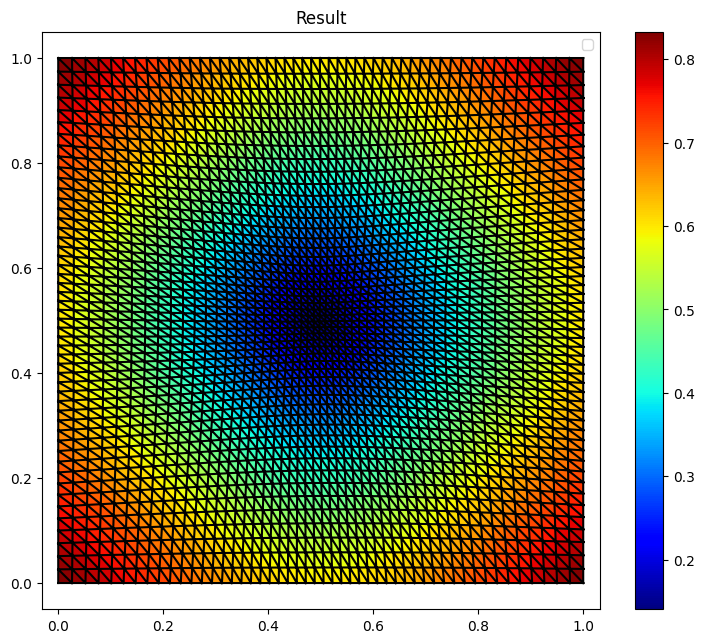}
        \caption{Dense Model Output}
    \end{subfigure}
    \caption{The experimental result obtained by our proposed LDEM method for the \emph{Ring-Shaped Population Distribution} test case.}
    \label{fig:ring_results}
\end{figure}

\begin{figure}[t!]
    \centering
    \begin{subfigure}[b]{0.31\textwidth}
        \includegraphics[width=\textwidth]{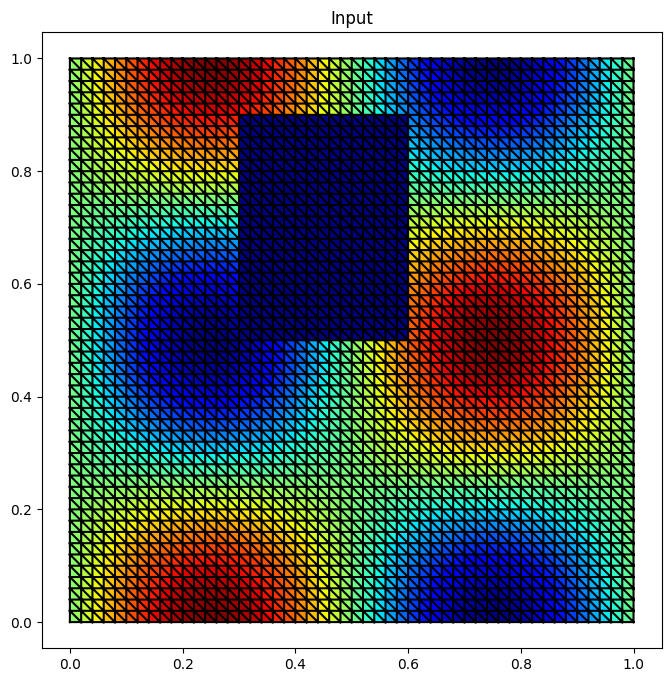}
        \caption{Input}
    \end{subfigure}
    \begin{subfigure}[b]{0.31\textwidth}
        \includegraphics[width=\textwidth]{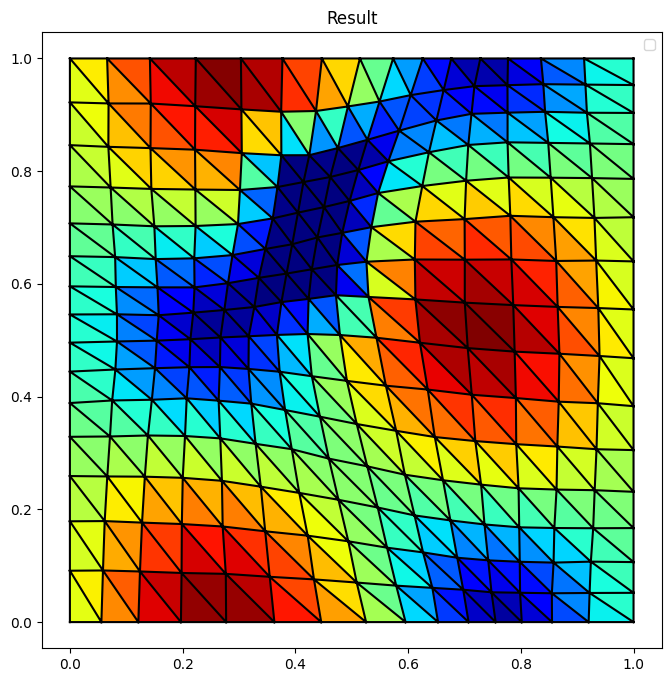}
        \caption{Coarse Model Output}
    \end{subfigure}
    \begin{subfigure}[b]{0.35\textwidth}
        \includegraphics[width=\textwidth]{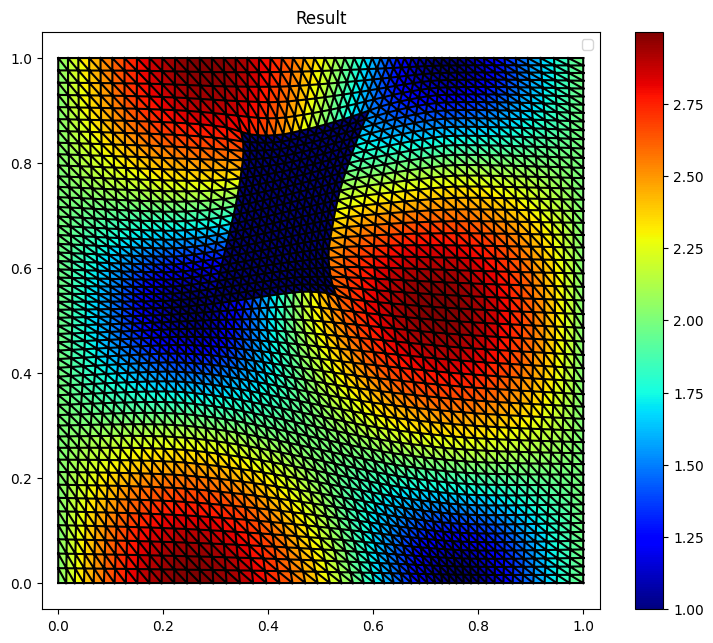}
        \caption{Dense Model Output}
    \end{subfigure}
    \caption{The experimental result obtained by our proposed LDEM method for the \emph{Localized Population Peaks} test case.}
    \label{fig:localized_peaks_results}
\end{figure}

\begin{figure}[t!]
    \centering

    \begin{subfigure}[b]{0.31\textwidth}
        \includegraphics[width=\textwidth]{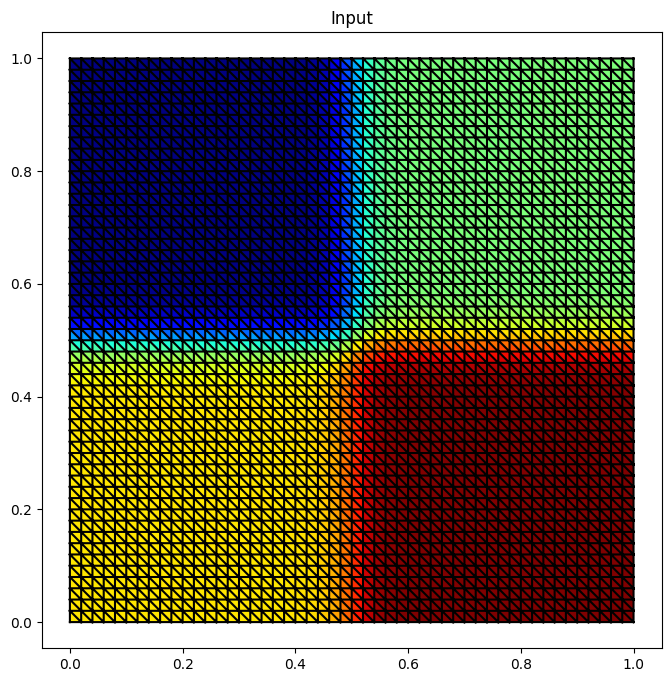}
        \caption{Input}
    \end{subfigure}
    \begin{subfigure}[b]{0.31\textwidth}
        \includegraphics[width=\textwidth]{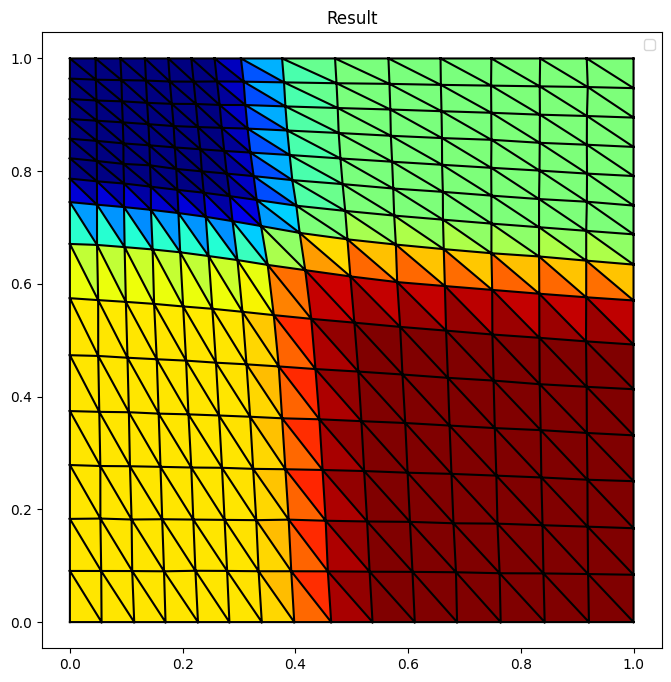}
        \caption{Coarse Model Output}
    \end{subfigure}
    \begin{subfigure}[b]{0.35\textwidth}
        \includegraphics[width=\textwidth]{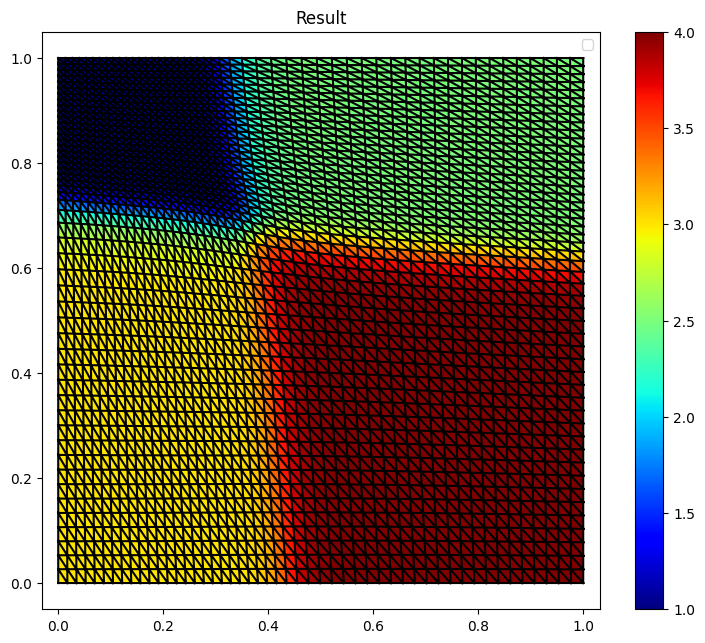}
        \caption{Dense Model Output}
    \end{subfigure}
    \caption{The experimental result obtained by our proposed LDEM method for the \emph{Smooth Blended Quadrants} test case.}
    \label{fig:blended_results}
\end{figure}

\begin{figure}[t!]
    \centering
    \begin{subfigure}[b]{0.31\textwidth}
        \includegraphics[width=\textwidth]{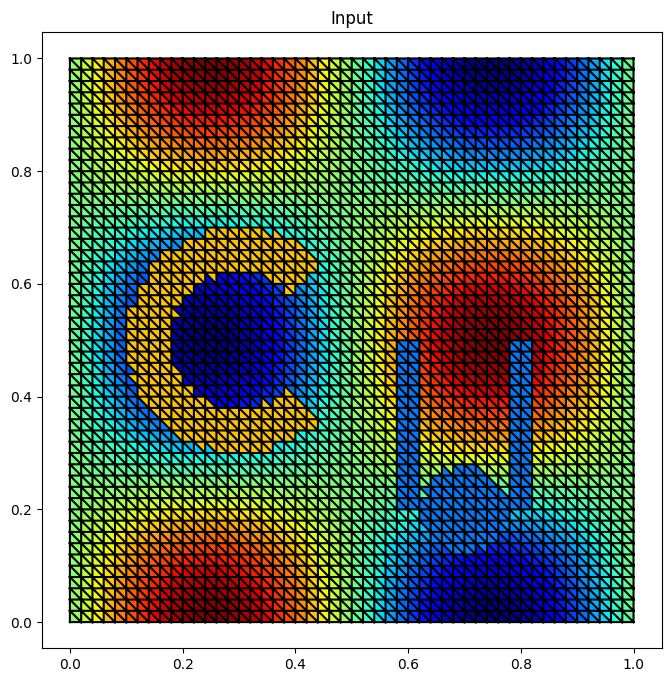}
        \caption{Input}
    \end{subfigure}
    \begin{subfigure}[b]{0.31\textwidth}
        \includegraphics[width=\textwidth]{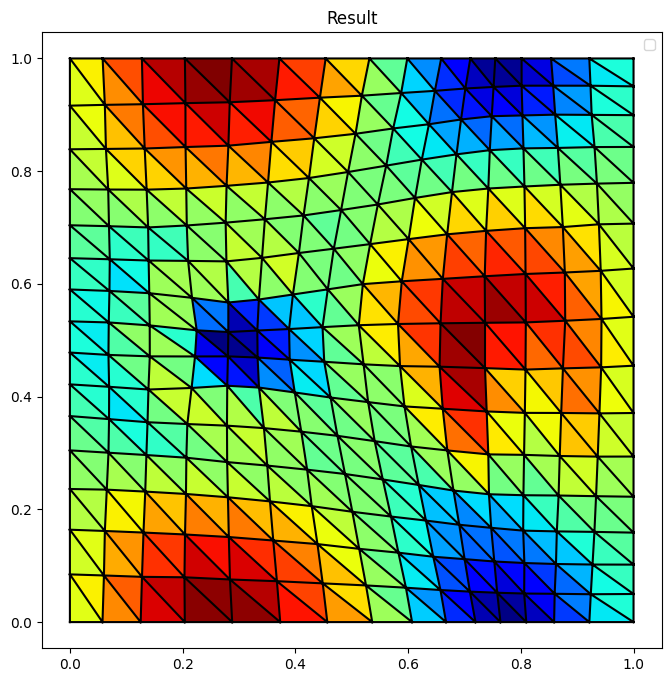}
        \caption{Coarse Model Output}
    \end{subfigure}
    \begin{subfigure}[b]{0.35\textwidth}
        \includegraphics[width=\textwidth]{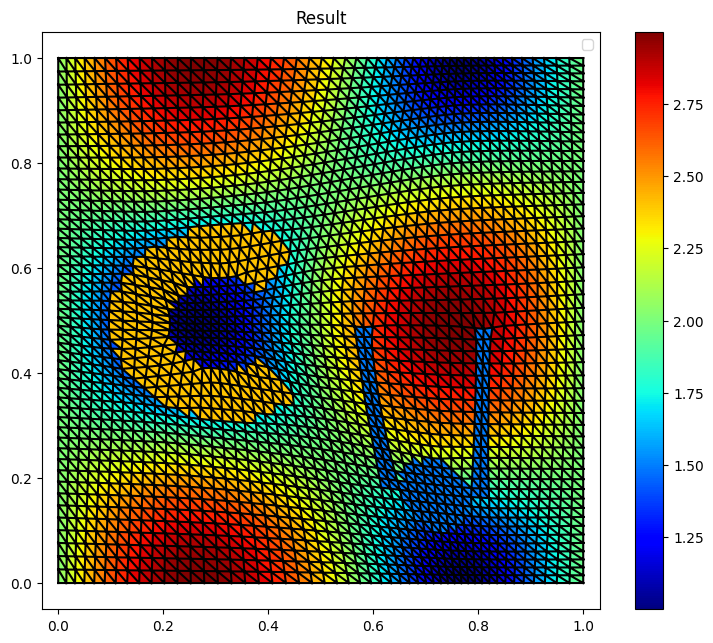}
        \caption{Dense Model Output}
    \end{subfigure}
    \caption{The experimental result obtained by our proposed LDEM method for the \emph{CU Pattern} test case.}
    \label{fig:cu_results}
\end{figure}

Below, we consider six scenarios (Fig.~\ref{fig:basic_sinusoidal_results}--\ref{fig:cu_results}) with a diverse set of population distributions to evaluate the model's ability to handle both simple and complex variations. The \emph{Input} shows the input population distribution on the dense grid, the \emph{Coarse Model Output} shows the intermediate output of the coarse model on the coarse grid, and the \emph{Dense Model Output} shows the final output of the dense model after refinement. From the color maps of the population distribution, we can have a qualitative assessment of the performance of the density-equalizing maps.

\begin{figure}[t!]
    \centering
    \begin{subfigure}[b]{0.31\textwidth}
        \includegraphics[width=\textwidth]{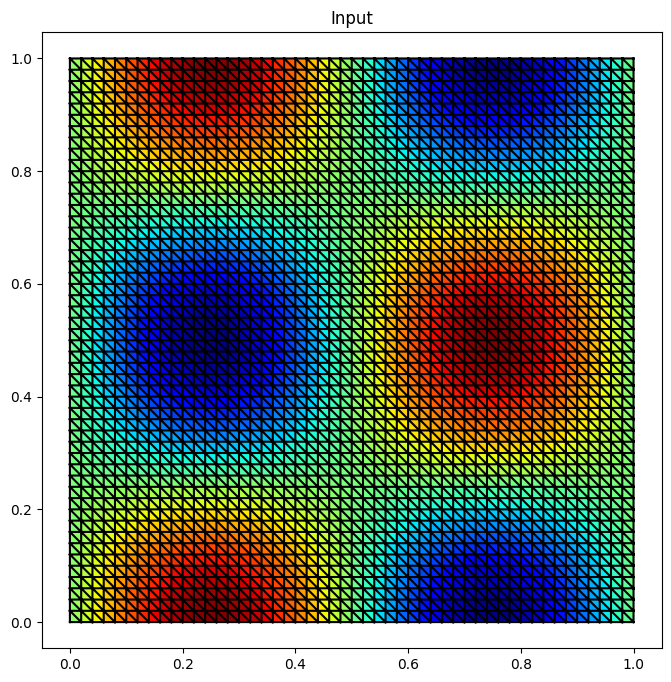}
        \caption{Input}
    \end{subfigure}
    \begin{subfigure}[b]{0.31\textwidth}
        \includegraphics[width=\textwidth]{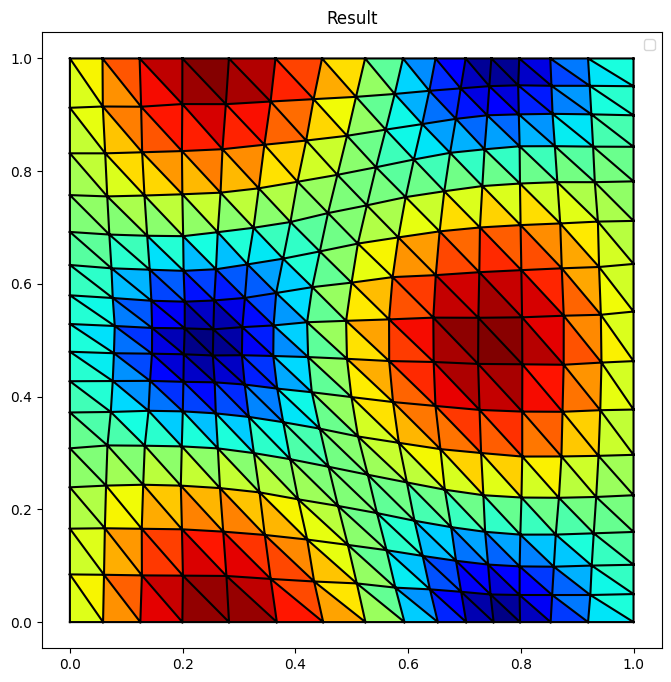}
        \caption{Coarse Model Output}
    \end{subfigure}
    \begin{subfigure}[b]{0.35\textwidth}
        \includegraphics[width=\textwidth]{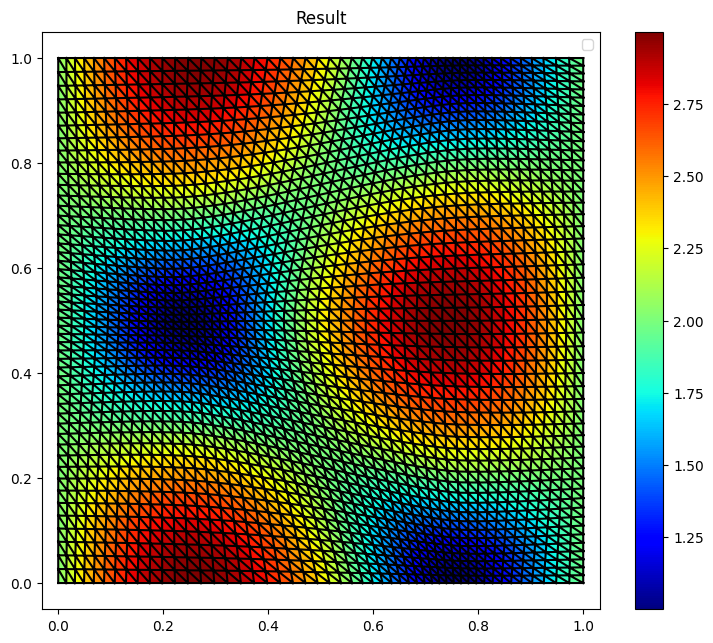}
        \caption{Dense Model Output}
    \end{subfigure}
    \caption{The experimental result obtained by our proposed LDEM method for the \emph{Basic Sinusoidal Variation} test case.}
    \label{fig:basic_sinusoidal_results}
\end{figure}

We start by considering an example of \emph{Basic Sinusoidal Variation} as shown in Fig.~\ref{fig:basic_sinusoidal_results}, with a smooth variation generated using sinusoidal functions:
\begin{equation}
\rho(\mathbf{c}) = 2 + \sin(2\pi c_x) \cdot \cos(2\pi c_y),
\end{equation}
where \(c_x\) and \(c_y\) are the \(x\)- and \(y\)-coordinates of the centroid $\mathbf{c}$. From the coarse and dense model outputs, we can easily see that regions with a lower population shrink and regions with a higher population expand. This suggests that the method can effectively produce area changes based on the prescribed population distributions.

Then, we consider an example of \emph{Complex Sinusoidal Variation} in Fig.~\ref{fig:complex_sinusoidal_results}. Specifically, to introduce additional complexity, exponential and logarithmic transformations are applied to the coordinates:
\begin{equation}
\rho(\mathbf{c}) = 2 + \sin\big(\exp(c_x) \cdot 2\pi\big) \cdot \cos\big(\log(c_y) \cdot \pi\big).
\end{equation}
In the mapping result, one can see that the method can successfully generate shape deformations with the desired area changes even for the nonlinear population variations with sharp transitions and unique patterns across the grid. 
        
Another example we consider is the \emph{Ring-Shaped Population Distribution} in Fig.~\ref{fig:ring_results}. Here, a ring-like population distribution is introduced, centered at \((0.5, 0.5)\) with a specified radius and thickness. The population decreases smoothly as the distance from the center of the ring deviates from the target radius. The distribution is defined as:
\begin{equation}
\rho(\mathbf{c}) = \exp\left(-\frac{(d(\mathbf{c}) - R)^2}{2 \cdot T^2}\right),
\end{equation}
where \(d(\mathbf{c}) = \sqrt{(c_x - 0.5)^2 + (c_y - 0.5)^2}\) is the distance from the centroid \(\mathbf{c}\) to the center of the ring, \(R\) is the radius of the ring and \(T\) is the thickness of the ring. This distribution creates a smooth circular band of population. From the density-equalizing mapping result, we can see that the central part is significantly shrunk while the four corner regions are enlarged.

Next, to simulate more complex scenarios, we consider the \emph{Localized Population Peaks} example in Fig.~\ref{fig:localized_peaks_results}. Here, specific rectangular regions within the grid are assigned higher population values. For a rectangle defined by \([x_{\min}, x_{\max}]\) and \([y_{\min}, y_{\max}]\), the population is updated as:
\begin{equation}
\rho(\mathbf{c}) = \begin{cases} 
P_{\text{peak}}, & \text{if } c_x \in [x_{\min}, x_{\max}] \text{ and } c_y \in [y_{\min}, y_{\max}], \\
\rho(\mathbf{c}), & \text{otherwise}.
\end{cases}
\end{equation}
It can be observed that even for such an input population with the sharp localized peaks, the proposed method can effectively create a density-equalizing map with different regions enlarged or shrunk respectively.

Another complex scenario we consider is the \emph{Smooth Blended Quadrants} as shown in Fig.~\ref{fig:blended_results}. To create smoother variations, a blend function is used to transition between different population levels across the grid quadrants. The smooth blending function is defined as:
\begin{equation}
S(x; c, w) = \frac{1}{1 + \exp\left(-\frac{x - c}{w}\right)},
\end{equation}
where \(c\) is the center, and \(w\) controls the blending width. The population is then defined for each quadrant:
\begin{equation}
\begin{aligned}
    \rho(\mathbf{c}) = & 1 \cdot \big(1 - S(c_x; 0.5, 0.02)\big) \cdot S(c_y; 0.5, 0.02) + \\
    & 2.5 \cdot S(c_x; 0.5, 0.02) \cdot S(c_y; 0.5, 0.02) + \\
    & 3 \cdot \big(1 - S(c_x; 0.5, 0.02)\big) \cdot \big(1 - S(c_y; 0.5, 0.02)\big) + \\
    & 4 \cdot S(c_x; 0.5, 0.02) \cdot \big(1 - S(c_y; 0.5, 0.02)\big).
\end{aligned}
\end{equation}
This gives an input population distribution consisting of four extreme regions with a smooth transition in between. From the mapping result, we can see that the regions are deformed smoothly, with the top-left region significantly shrunk and the bottom-right region enlarged.

Finally, we consider the \emph{Complex Patterns} example in Fig.~\ref{fig:cu_results}. Here, the population distribution is further extended to include predefined patterns, such as the characters ``CU''. A binary mask is used to define the locations of the pattern, and the population values within these regions are set to higher values to simulate localized emphasis:
\begin{equation}
\rho(\mathbf{c}) = \rho_{\text{base}} + \Delta_{\text{pattern}}, \quad \text{for centroids within the pattern}.
\end{equation}
Even for such a complex input population distribution, it can be observed in the final output that the shape deformation satisfies the desired effect very well.

\begin{figure}[t!]
    \centering
    \includegraphics[width=0.75\textwidth]{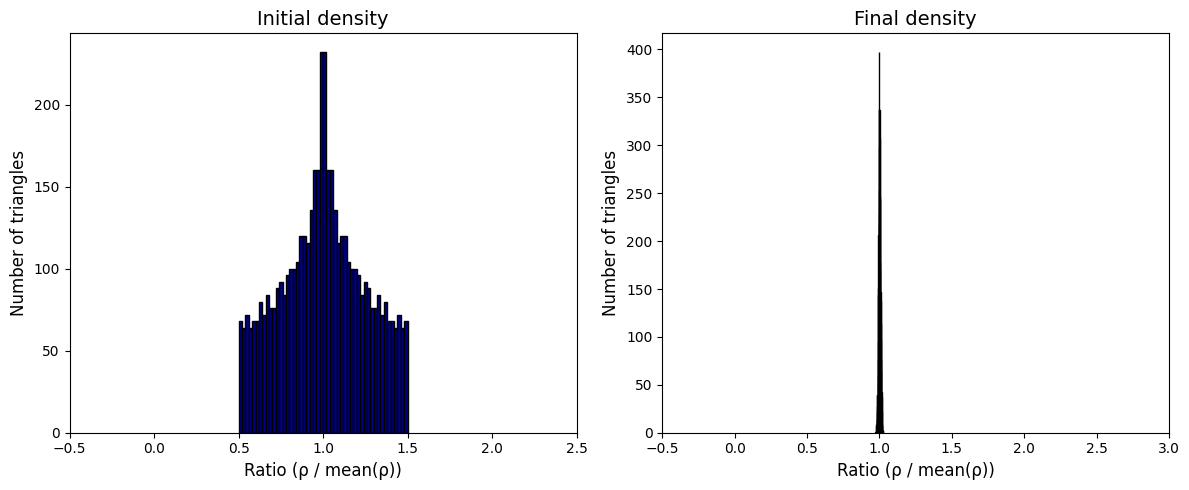}
    \caption{The histograms of the density distribution for the \emph{Basic Sinusoidal Variation} test case.}
    \label{fig:basic_distribution}
\end{figure}

\begin{figure}[t!]
    \centering
    \includegraphics[width=0.75\textwidth]{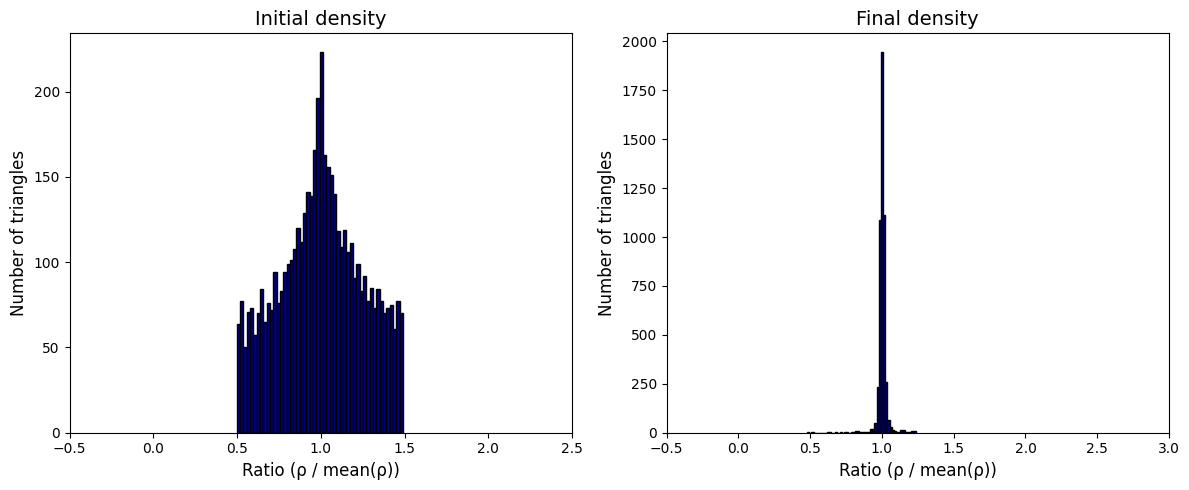}
    \caption{The histograms of the density distribution for the \emph{Complex Sinusoidal Variation} test case.}
    \label{fig:complex_distribution}
\end{figure}

\begin{figure}[t!]
    \centering
    \includegraphics[width=0.75\textwidth]{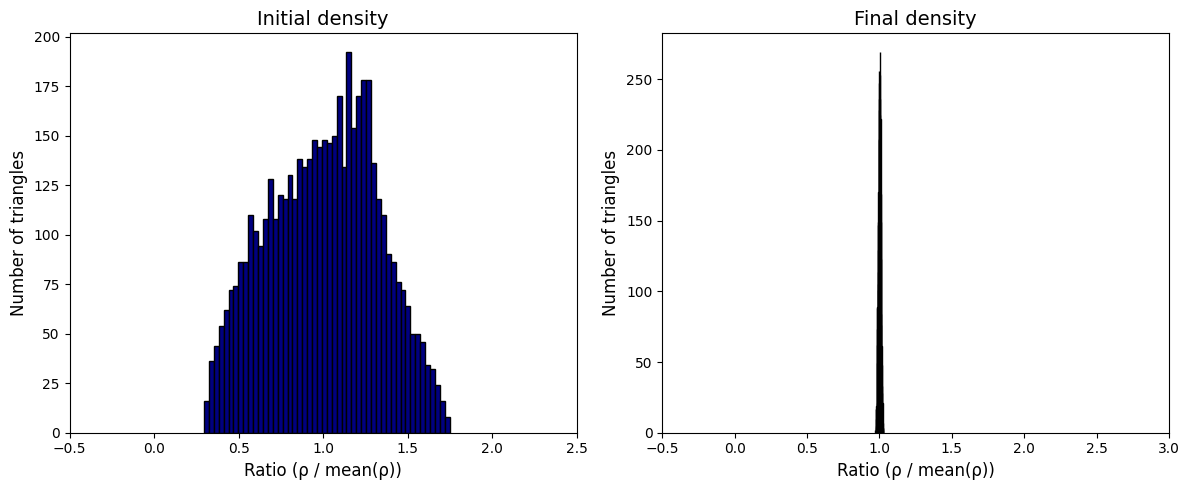}
    \caption{The histograms of the density distribution for the \emph{Ring-Shaped Population Distribution} test case.}
    \label{fig:ring_distribution}
\end{figure}

\begin{figure}[t!]
    \centering
    \includegraphics[width=0.75\textwidth]{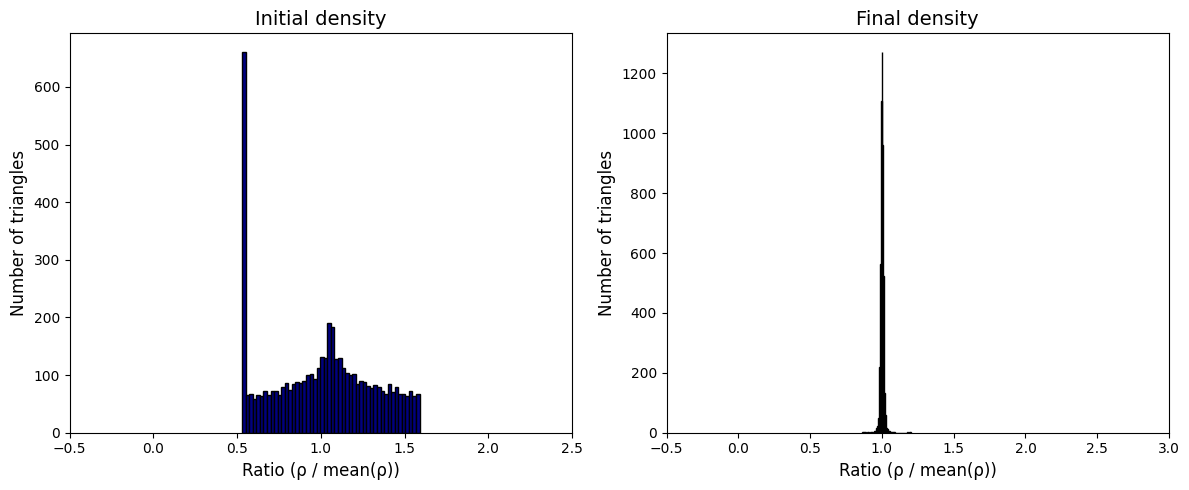}
    \caption{The histograms of the density distribution for the \emph{Localized Population Peaks} test case.}
    \label{fig:peak_distribution}
\end{figure}

\begin{figure}[t!]
    \centering
    \includegraphics[width=0.75\textwidth]{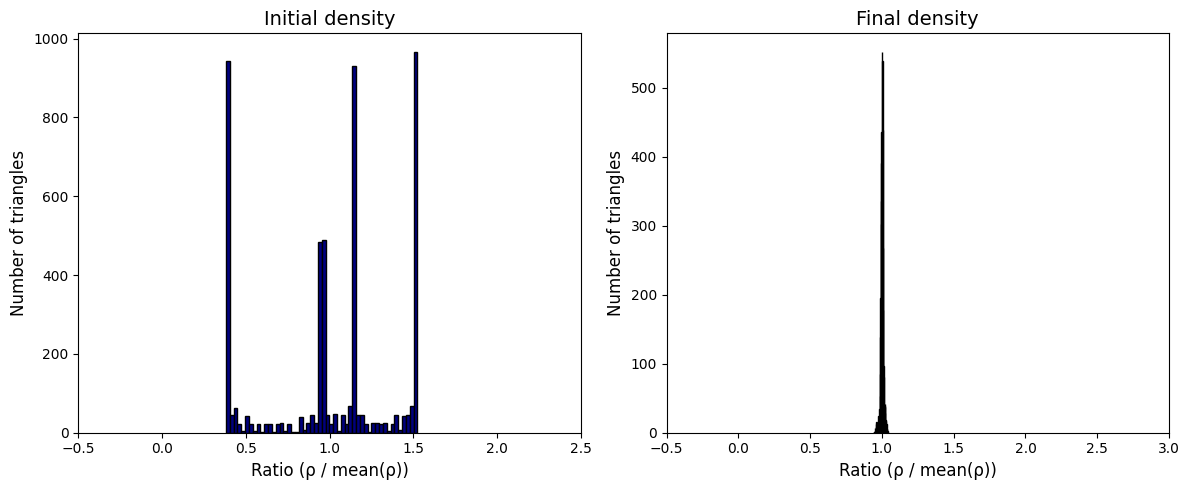}
    \caption{The histograms of the density distribution for the \emph{Smooth Blended Quadrants} test case.}
    \label{fig:blend_distribution}
\end{figure}

\begin{figure}[t!]
    \centering
    \includegraphics[width=0.75\textwidth]{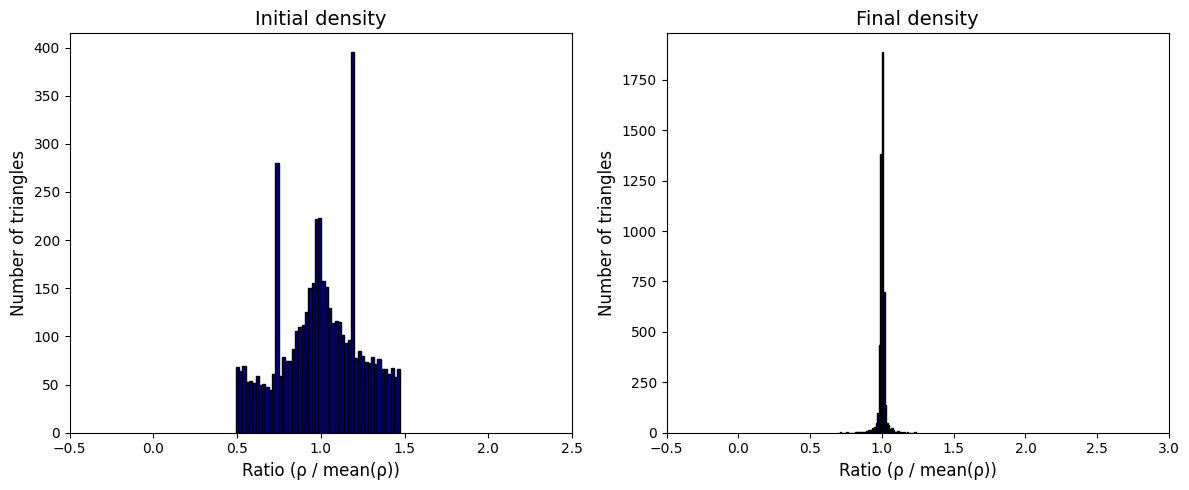}
    \caption{The histograms of the density distribution for the \emph{CU Pattern} test case.}
    \label{fig:cu_distribution}
\end{figure}

The above examples demonstrate the effectiveness of our proposed method for generating density-equalizing maps with a large variety of desired effects. Next, for a more quantitative assessment of the quality of the mappings, we compute the density distribution using the given population and the initial or final mapping result:
\begin{equation}
\rho = \frac{\text{Population on face}}{\text{Area of face}}.
\end{equation}
In Fig.~\ref{fig:basic_distribution}--\ref{fig:cu_distribution}, we plot the histograms of the initial density ($\frac{\text{Population}}{\text{Initial area}}$) and final density ($\frac{\text{Population}}{\text{Final area}}$) for all mapping examples. For a better visualization, we further divide $\rho$ by $\text{mean}(\rho)$ in the histograms. It can be observed that for all examples, the final density distribution is highly concentrated and significantly sharper than the initial density distribution. This indicates that the mapping results produced by our proposed method are highly density-equalizing.

\subsection{Comparison with the traditional diffusion-based iterative method}
In Table~\ref{table:error}, we further compare our proposed DNN-based method with the traditional diffusion-based iterative method~\cite{choi2018density}, which requires iteratively solving the diffusion equation and updating the vertex positions. First, we define the \emph{Density-Equalizing Error} (DE Error) as the ratio std($\rho$)/mean($\rho$) and use it to quantify how uniformly the population is distributed in the mapping result. It can be observed that the proposed method achieves comparable DE Error values with the traditional iterative method in all examples. Moreover, for the examples with sharp or complex population transitions (Fig.~\ref{fig:complex_sinusoidal_results}, ~\ref{fig:localized_peaks_results}, and~\ref{fig:cu_results}), our method can significantly reduce the DE Error by over 30\% when compared to the traditional method. We also assess the conformal distortion and the bijectivity of the mappings using the Beltrami coefficient. Specifically, the average norm of the Beltrami coefficient of the mapping, denoted as BC-mean, gives a measure of the conformal distortion. It can be observed that the BC-mean values achieved by our method are all close to 0 and comparable to those obtained from the traditional method. This suggests that the local geometric distortion of the mappings is small. Also, the maximum value of the norm of the Beltrami coefficient (BC-max) is less than 1 for all examples, which indicates that the mappings are bijective. 

\begin{table}[t!]
    \centering
    \begin{tabular}{@{}lccc|ccc@{}}
        \toprule
        \textbf{Test Case} & \multicolumn{3}{c|}{\textbf{Our proposed LDEM method}} & \multicolumn{3}{c}{\textbf{Diffusion-based method~\cite{choi2018density}}} \\
        \cmidrule(lr){2-4} \cmidrule(l){5-7}
         & BC-mean & BC-max & DE Error & BC-mean & BC-max & DE Error \\
        \midrule
        Fig.~\ref{fig:basic_sinusoidal_results}(a) & 0.1144 & 0.2732 & 0.0069   & 0.1109 & 0.2600 & 0.0057 \\
        Fig.~\ref{fig:complex_sinusoidal_results}(a) & 0.1136 & 0.3836 & 0.0436   & 0.1125 & 0.3071 & 0.0733 \\
        Fig.~\ref{fig:ring_results}(a) & 0.1479 & 0.2786 & 0.0084   & 0.1451 & 0.3146 & 0.0071 \\
        Fig.~\ref{fig:localized_peaks_results}(a) & 0.1390 & 0.3964 & 0.0127   & 0.1413 & 0.4389 & 0.0246 \\
        Fig.~\ref{fig:blended_results}(a) & 0.1847 & 0.3844 & 0.0102   & 0.1819 & 0.3564 & 0.0088 \\
        Fig.~\ref{fig:cu_results}(a) & 0.1059 & 0.3512 & 0.0233   & 0.1042 & 0.3203 & 0.0340 \\
        \bottomrule
    \end{tabular}
    \caption{Comparison of the proposed DNN method and the diffusion-based iterative method~\cite{choi2018density} for six test cases, including \emph{Basic Sinusoidal Variation} (Fig.~\ref{fig:basic_sinusoidal_results}(a)), \emph{Complex Sinusoidal Variation} (Fig.~\ref{fig:complex_sinusoidal_results}(a)), \emph{Ring-Shaped Population Distribution} (Fig.~\ref{fig:ring_results}(a)), \emph{Localized Population Peaks} (Fig.~\ref{fig:localized_peaks_results}(a)),  \emph{Smooth Blended Quadrants} (Fig.~\ref{fig:blended_results}(a)), and \emph{CU Pattern} (Fig.~\ref{fig:cu_results}(a)). Here, BC-mean represents the average norm of the Beltrami coefficient $|\mu|$, BC-max represents the maximum value of the norm of the Beltrami coefficient $|\mu|$, and the DE Error represents the density-equalizing error, defined as the ratio std($\rho$)/mean($\rho$).}
    \label{table:error}
\end{table}

\begin{figure}[t!]
    \centering
    \begin{subfigure}[b]{0.31\textwidth}
        \includegraphics[width=\textwidth]{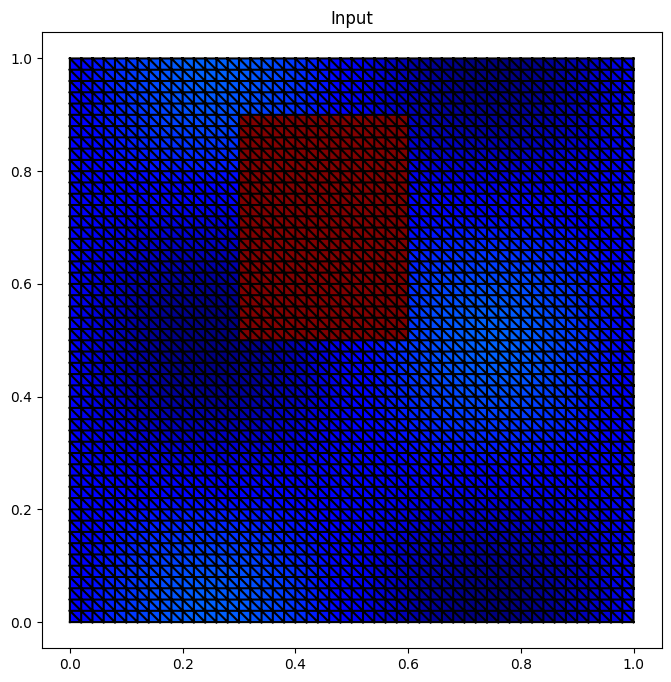}
        \caption{Input}
    \end{subfigure}
    \begin{subfigure}[b]{0.31\textwidth}
        \includegraphics[width=\textwidth]{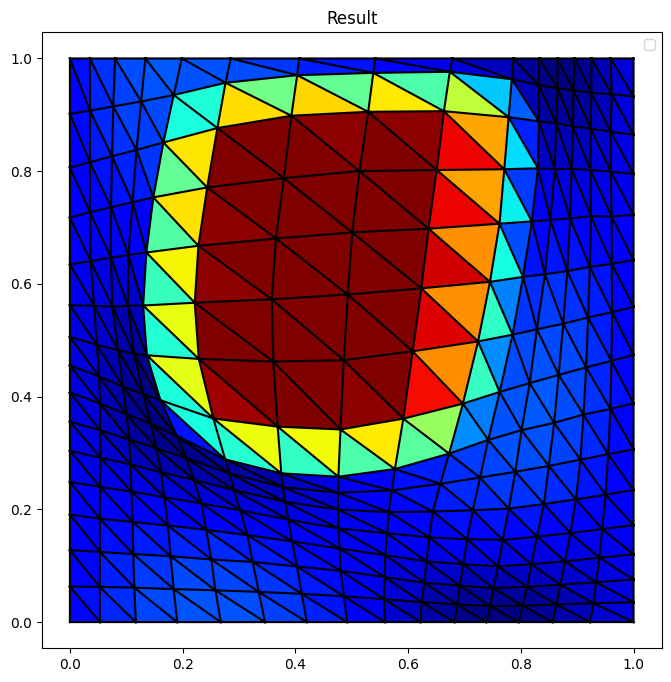}
        \caption{Coarse Model Output}
    \end{subfigure}
    \begin{subfigure}[b]{0.35\textwidth}
        \includegraphics[width=\textwidth]{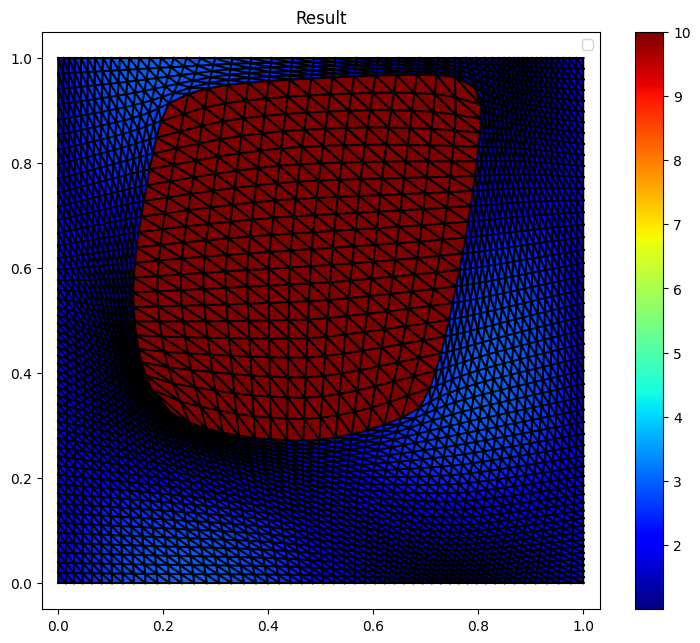}
        \caption{Dense Model Output}
    \end{subfigure}
    \caption{The experimental result obtained by our proposed method for the \emph{Extreme} test case.}
    \label{fig:extreme_results}
\end{figure}

Next, we further consider a test case with \emph{Extreme} populations as shown in Fig.~\ref{fig:extreme_results}(a). Specifically, we consider a prescribed population distribution with an extremely large value of 10 inside a specific rectangular region, and much smaller values outside the rectangular region. From Fig.~\ref{fig:extreme_results}(b)--(c), it can be observed that our proposed method can handle such an extreme case very well, resulting in a large shape deformation without mesh overlaps.

\begin{figure}[t!]
    \centering
    \begin{subfigure}[b]{0.31\textwidth}
        \includegraphics[width=\textwidth]{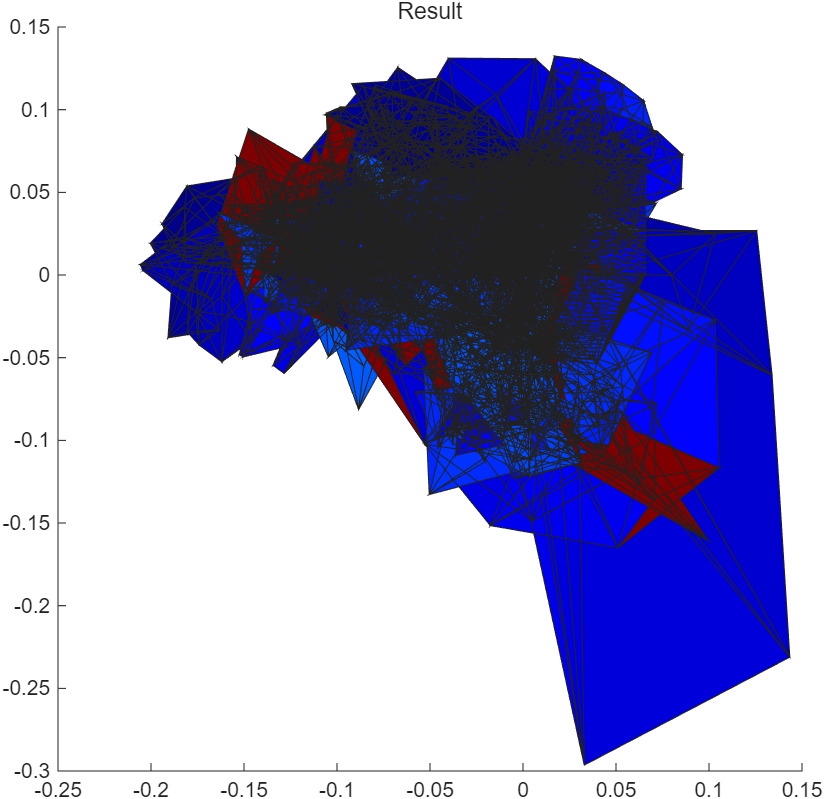}
        \caption{Diffusion (Default step size)}
    \end{subfigure}
    \begin{subfigure}[b]{0.31\textwidth}
        \includegraphics[width=\textwidth]{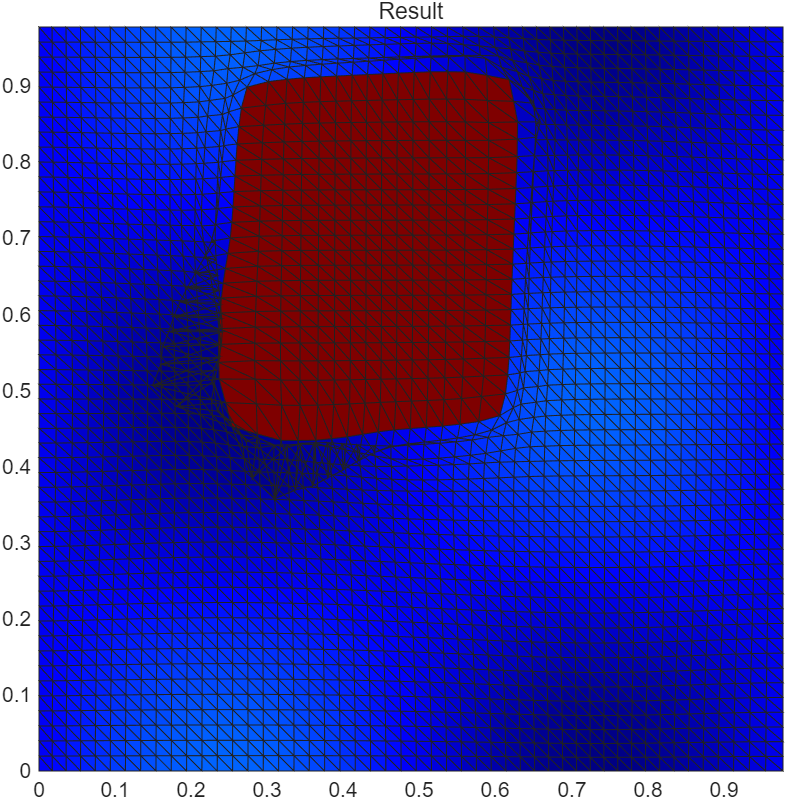}
        \caption{Diffusion (Reduced step size)}
    \end{subfigure}
    \begin{subfigure}[b]{0.35\textwidth}
        \includegraphics[width=\textwidth]{figures/extreme_dense_output.png}
        \caption{Our Method}
    \end{subfigure}
    \caption{The mapping results obtained by the traditional diffusion-based iterative method~\cite{choi2018density} with different step sizes and our proposed method for the \emph{Extreme} Test Case.}
    \label{fig:extreme_comparison}
\end{figure}

On the contrary, for the traditional diffusion-based iterative method, using the default step size (defined by Eq.~(4.39) in~\cite{choi2018density}) will lead to severe mesh overlaps (Fig.~\ref{fig:extreme_comparison}(a)). A possible explanation is that the default step size $\delta t$ in~\cite{choi2018density} is dependent on the distribution of the initial density $\rho$, and for this extreme test case, $\delta t$ becomes very large and hence the algorithm produces severe overlaps. We also consider reducing the step size to a much smaller value ($\delta t = 5 \times 10^{-5}$) in the diffusion-based method to alleviate the overlapping issue, but that leads to an inaccurate mapping result in which the high-density region is not sufficiently enlarged in the final mapping result (Fig.~\ref{fig:extreme_comparison}(b)). In Table~\ref{table:error_extreme}, we further compare the mapping results obtained by different approaches in terms of the density-equalizing error (DE Error), local geometric distortion (BC-mean), and bijectivity measure (BC-max). It is easy to see that the diffusion-based method with the default step size gives a highly inaccurate and non-bijective result. While reducing the step size can alleviate the overlapping issue, the result is still non-bijective, and the density-equalizing error remains large. By contrast, our proposed method can significantly reduce the density-equalizing error by over 98\% and preserve the bijectivity.

\begin{table}[t]
    \centering
    \begin{tabular}{@{}lccc@{}}
        \toprule
        \textbf{Method} & \textbf{BC-mean} & \textbf{BC-max} & \textbf{DE Error} \\
        \midrule
        Diffusion with default step size (Fig.~\ref{fig:extreme_comparison}(a)) & 1.4108 & 25.8513 & 46.6316 \\
        Diffusion with reduced step size (Fig.~\ref{fig:extreme_comparison}(b)) & 0.1127 & 1.8490 & 2.6494 \\
        Our proposed LDEM method (Fig.~\ref{fig:extreme_comparison}(c)) & 0.3028 & 0.7778 & 0.0371 \\
        \bottomrule
    \end{tabular}
    \caption{Quantitative comparison between the traditional diffusion-based iterative method~\cite{choi2018density} and our proposed method for the \emph{Extreme} Test Case.}
    \label{table:error_extreme}
\end{table}

Altogether, the experimental results show that when compared to the traditional diffusion-based methods, our proposed LDEM method can effectively handle a wider range of population distributions and produce highly accurate density-equalizing maps, while maintaining comparably low geometric distortion and preserving the bijectivity.

\section{Application to surface remeshing}
\label{sect:application}
In engineering and graphics, it is common to remesh surfaces to control the triangulation density and quality. This is important for many applications such as shape modelling, solving PDEs on surfaces, and visualization. Using the proposed LDEM method, we can easily perform surface remeshing with different desired effects.

More specifically, given a simply-connected open triangulated surface $\mathcal{M}$, we can first follow the procedure in~\cite{choi2018density} and parameterize it onto a square domain using the Tutte embedding method. Denote this initial mapping as $g: \mathcal{M} \to [0,1]^2$. Then, we can easily compute a density-equalizing map $f: [0,1]^2 \to [0,1]^2$ on the unit square using the proposed LDEM method, with the prescribed population controlling the desired effect. Here, for regions that we desire to have a denser triangulation, a higher population can be set. Conversely, for regions that we desire to have a coarser triangulation, a lower population can be set. Then, under the LDEM mapping $f$, different regions will be enlarged or shrunk accordingly. After getting the mapping result, we can use the inverse mapping $(f \circ g)^{-1}$ to map a uniform triangulation generated on the unit square back to the original surface $\mathcal{M}$. Because of the shape deformation achieved by LDEM, the new triangulation on $\mathcal{M}$ will generally become non-uniform, with different regions having higher or lower mesh densities following the prescribed effect. This completes the surface remeshing process.

\begin{figure}[t!]
    \centering
    \includegraphics[width=\linewidth]{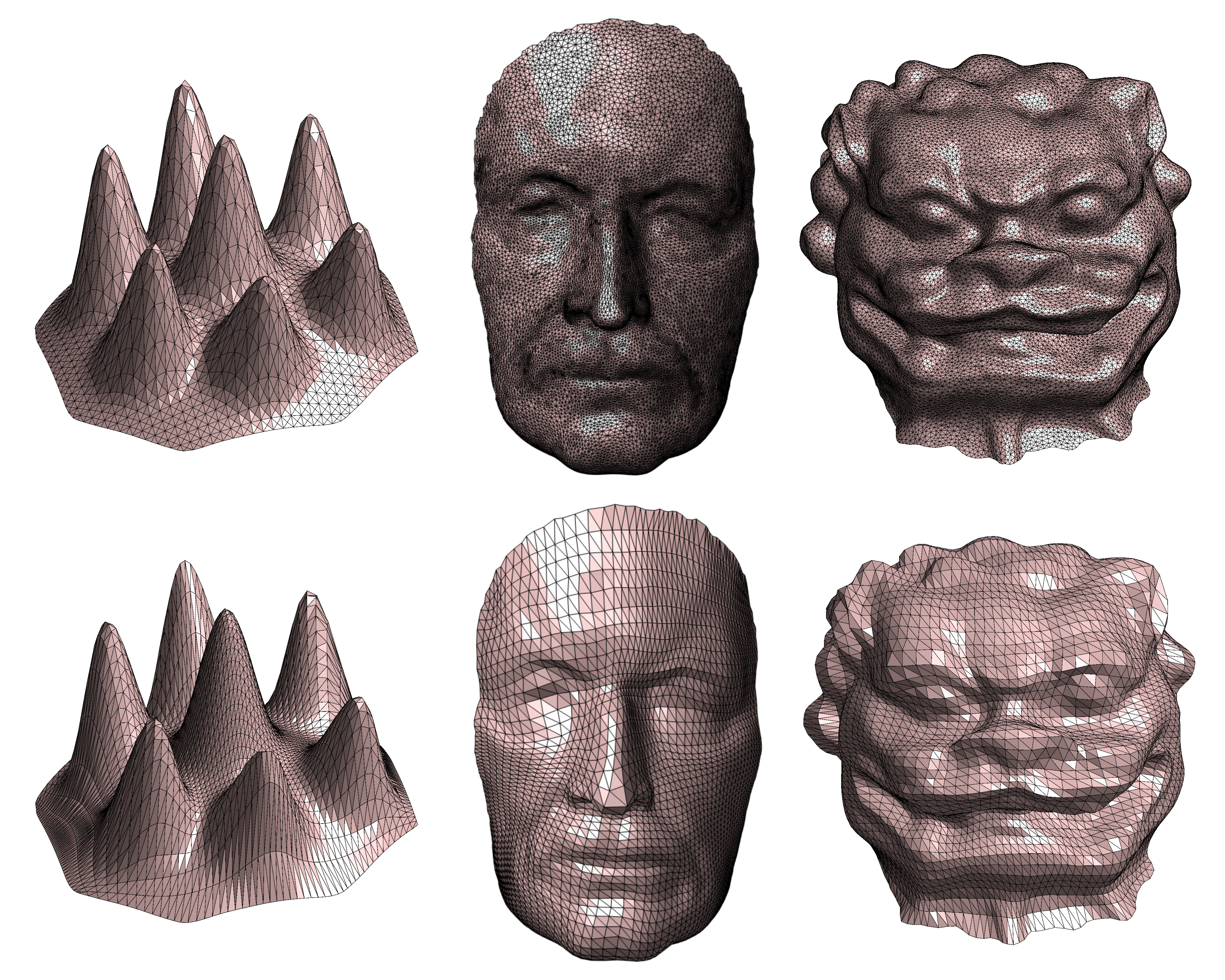}
    \caption{Surface remeshing achieved by our proposed LDEM method. Left to right: \emph{Peaks}, \emph{Max Planck}, and \emph{Chinese Lion}. The top row shows the original surfaces, and the bottom row shows the remeshed surfaces with different effects.}
    \label{fig:remeshing}
\end{figure}

Fig.~\ref{fig:remeshing} shows three sets of surface remeshing examples with different effects achieved by our method. For the \emph{Peaks} example, which is a surface with multiple peaks, we would like to have a higher mesh resolution at the central peak in the remeshing result. To achieve this, we set the population in the central region to be much larger than that in the other regions and apply the above-mentioned approach. It can be observed that because of the density-equalizing property of our method, the remeshed surface achieves the desired effect very well. In the \emph{Max Planck} example, we set the population in the top left region of the surface to be lower than that in the top right region. Consequently, the mesh density in the top left region is lower in the remeshing result. Alternatively, we can also achieve a relatively uniform remeshing result, as shown in the \emph{Chinese Lion} example, by adjusting the population in different regions. Besides, as mentioned in the previous section, our proposed method preserves the bijectivity very well. Here in the surface remeshing experiments, it can also be observed that the remeshed surfaces are all folding-free. 

Altogether, the experiments demonstrate the advantage of our proposed method for surface remeshing for engineering applications.

\section{Extension to 3D} \label{sect:3d}
Our proposed LDEM method can be naturally extended to 3D. In particular, note that the extension to 3D (denoted as LDEM-3D) does not require modifying the general model structure. Instead, we will only need to minimally adjust the training data and the loss function for the 3D case.

\begin{figure}[t!]
    \centering
    \begin{subfigure}[t]{0.48\linewidth}
        \centering
        \includegraphics[width=\linewidth]{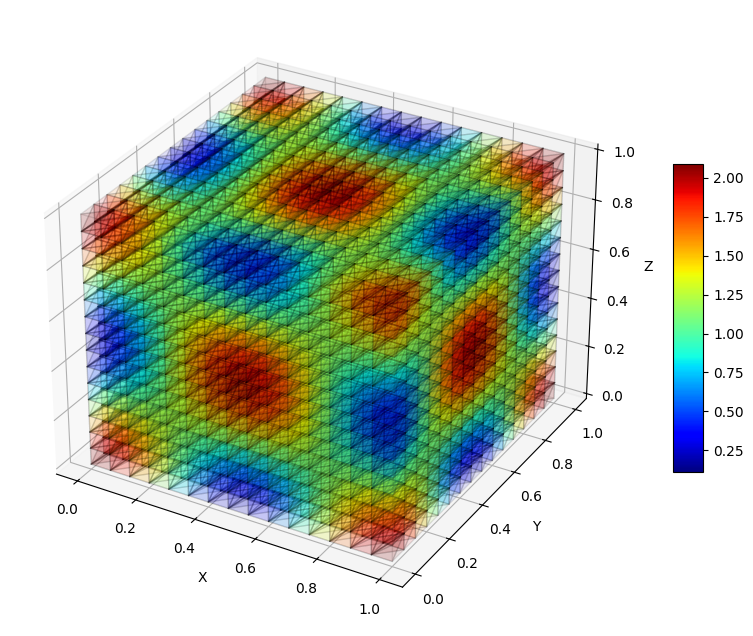}
        \caption{Initial domain}
        \label{fig:initial_map_3d}
    \end{subfigure}
    \hfill
    \begin{subfigure}[t]{0.48\linewidth}
        \centering
        \includegraphics[width=\linewidth]{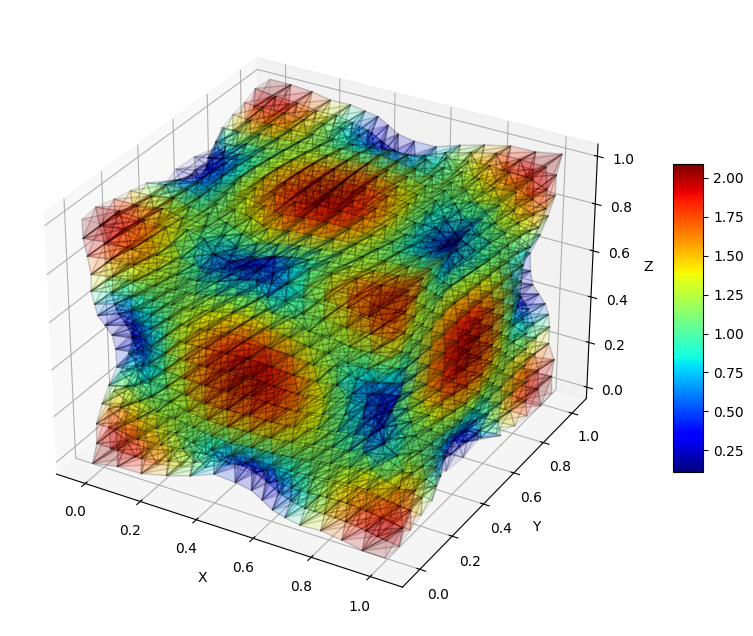}
        \caption{Final 3D density-equalizing mapping result}
        \label{fig:transformed_map_3d}
    \end{subfigure}
    \caption{An illustration of the 3D density-equalizing maps. Both the initial domain and final mapping result are tetrahedralized and color-coded with the given population. The transformation of the initial tetrahedralized domain adjusts the vertex positions so that the tetrahedron volumes match the target density values.}
    \label{fig:initial_and_transformed_maps_3d}
\end{figure}

\subsection{Model Formulation}

Extending the LDEM approach to 3D naturally follows the methodology developed for the 2D case. The primary objective remains to transform an initial 3D domain into one with uniform population density, achieved by adjusting local volumes proportional to assigned population values. To achieve this, we initialize the domain using a regular $N \times N \times N$ meshgrid. A tetrahedralization is applied to divide the domain into non-overlapping tetrahedra, each associated with a given population value. The 3D density-equalizing map aims to relocate the mesh points such that the resulting map equalizes the density, defined as the population per unit volume (see Fig.~\ref{fig:initial_and_transformed_maps_3d} for an illustration).

More specifically, consider a 3D domain discretized into tetrahedra. Each tetrahedron $T_i$ has an initial population value $p_i$. The density equalization aims to produce a new configuration where the volume of each transformed tetrahedron $T'_i$ satisfies:
\begin{equation}
\text{vol}(T'_i) \propto p_i.
\end{equation}

Analogous to the 2D case, here we can consider a coarse model with the aid of a coarse 3D grid. The grid is constructed with \(D_{\text{coarse}} \times D_{\text{coarse}} \times D_{\text{coarse}}\) vertices uniformly distributed over the unit cube \([0, 1]^3\). The vertex coordinates are given by:
\begin{equation}
    x_i, y_j, z_k \in \left\{ 0, \frac{1}{D_{\text{coarse}}-1}, \frac{2}{D_{\text{coarse}}-1}, \ldots, 1 \right\}, \quad i, j, k = 1, 2, \ldots, D_{\text{coarse}}. 
\end{equation}
Tetrahedral elements are defined by connecting adjacent vertices to form multiple tetrahedra within each cubic cell. This ensures that the dense 3D grid is fully tetrahedralized. The population distribution is modeled as a continuous function over the centroids of tetrahedral elements. For a tetrahedron with vertices \(\mathbf{v}_a, \mathbf{v}_b, \mathbf{v}_c, \mathbf{v}_d\), the centroid is given by:
\begin{equation}
    \mathbf{c} = \frac{\mathbf{v}_a + \mathbf{v}_b + \mathbf{v}_c + \mathbf{v}_d}{4}.
\end{equation}
A population value $p_i$ can then be defined at every centroid $\mathbf{c}_i$.

To quantify the density-equalizing effect, we define the density uniformity loss analogous to the 2D case, but now employing volume measures:
\begin{equation}
    \mathcal{L}_{\text{density3D}} = \frac{\mathrm{std}(\boldsymbol{\rho}_T)}{\mathrm{mean}(\boldsymbol{\rho}_T)}, 
\end{equation}
where $\boldsymbol{\rho}_T$ is the vector of tetrahedron-wise densities defined by
\begin{equation}
     \rho_{T,i} = \frac{p_i}{\text{vol}(T_i')}.
\end{equation}

For geometric regularization, the loss function $\mathcal{L}_{\text{distance}}$ in Eq.~\eqref{eqt:distance} can be extended naturally to distances in $\mathbb{R}^3$:
\begin{equation}
\mathcal{L}_{\text{distance3D}} = \frac{1}{N}\sum_{i=1}^{N}\left(\text{distance}_{x,i} + \text{distance}_{y,i} + \text{distance}_{z,i}\right),
\end{equation}
where $\text{distance}_{x,i}$, $\text{distance}_{y,i}$, $\text{distance}_{z,i}$ are defined based on the squared distance between neighboring points in the $x$-, $y$-, and $z$-axis groups, respectively. Note that one could also generalize the $\mathcal{L}_{\text{slope}}$ in Eq.~\eqref{eqt:slope} to 3D, but the consideration of slope in 3D involves several more terms. For simplicity, we omit this function in the subsequent discussion and experiments. 

Altogether, we have the following overall loss function for training in the 3D case:
\begin{equation}
\mathcal{L}_{\text{3D}} = \lambda_d \cdot \mathcal{L}_{\text{density3D}} + \lambda_l \cdot \mathcal{L}_{\text{distance3D}}.
\end{equation}
By following this formulation, the model directly generalizes to three-dimensional scenarios without altering its core architecture, requiring only adjustments to the input data representation and dimensionality of computations.

For the hyperparameters, we have the following settings:
\begin{itemize}
    \item For the initialization phase, we set the learning rate as \( \text{init\_lr\_coarse} = 1 \times 10^{-2} \) and the number of epochs as \( \text{init\_epochs\_coarse} = 1500 \). This stage provides a well-conditioned starting point for later optimization.
    
    \item For the training phase, we set the learning rate as \( \text{train\_lr\_coarse} = 1 \times 10^{-4} \) and the maximum number of epochs as \( \text{max\_epochs\_coarse} = 5000 \). The early stopping patience is set as 200 epochs. The minimum improvement threshold is \( \text{min\_delta} = 1 \times 10^{-4} \), and the warm-up period is 150 epochs (during which early stopping is disabled). Early stopping is used to prevent overfitting, and the warm-up period helps stabilize training before convergence is monitored.
\end{itemize}

All other parts of the 2D framework, including the coarse-to-dense transformation and fine-tuning with dense model, can be extended to 3D in a similar manner.

\subsection{Experimental results}
To simplify the experiment process, we only test the coarse model for the 3D extension as we could deal with the dense case similarly as we did in the 2D cases. Below, we set \(D_{\text{coarse}} = 16\) and \(\lambda_d = \lambda_l = 1\). Different population distributions are designed to capture various scenarios, ranging from simple to complex spatial variations.

\begin{figure}[t!]
    \centering
    \begin{subfigure}[t]{0.48\linewidth}
        \centering
        \includegraphics[width=\linewidth]{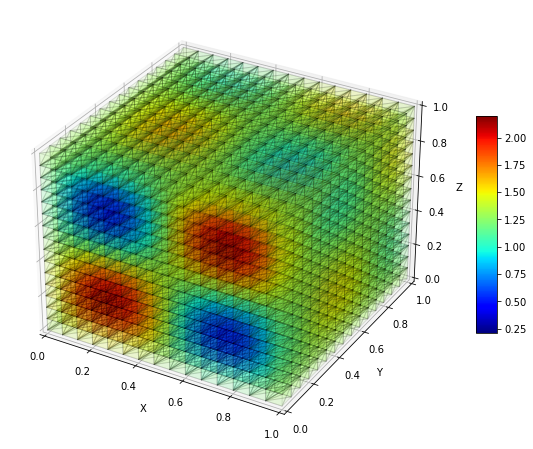}
        \caption{Input}
    \end{subfigure}
    \hfill
    \begin{subfigure}[t]{0.48\linewidth}
        \centering
        \includegraphics[width=\linewidth]{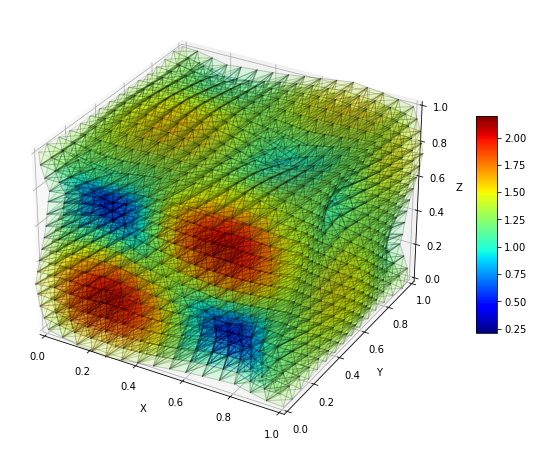}
        \caption{Output}
    \end{subfigure}
    \caption{The experimental result obtained by our proposed LDEM-3D method for the \emph{3D Basic Sinusoidal Variation} test case.}
    \label{fig:basic_sinusoidal_3d}
\end{figure}

\begin{figure}[t!]
    \centering
    \begin{subfigure}[t]{0.48\linewidth}
        \centering
        \includegraphics[width=\linewidth]{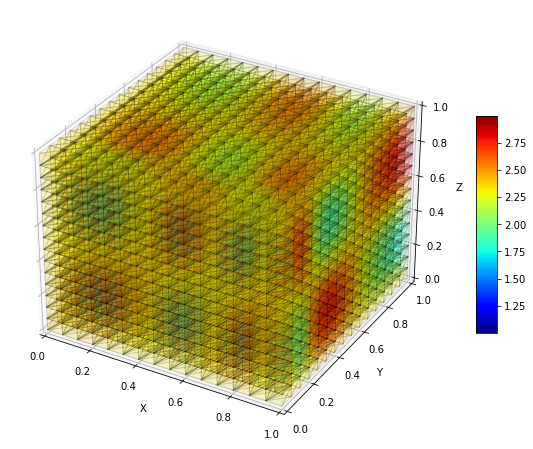}
        \caption{Input}
    \end{subfigure}
    \hfill
    \begin{subfigure}[t]{0.48\linewidth}
        \centering
        \includegraphics[width=\linewidth]{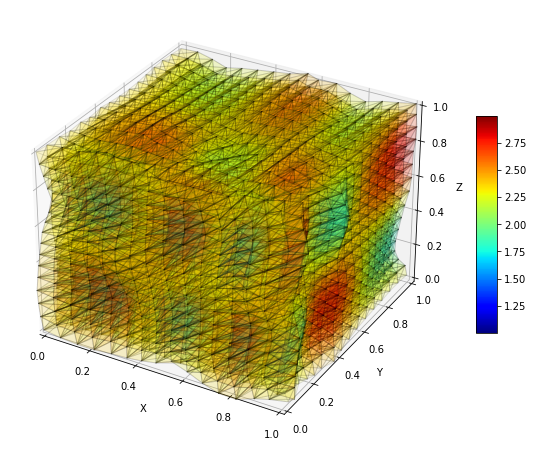}
        \caption{Output}
    \end{subfigure}
    \caption{The experimental result obtained by our proposed LDEM-3D method for the \emph{3D Complex Sinusoidal Variation} test case.}
    \label{fig:complex_sinusoidal_3d}
\end{figure}

\begin{figure}[t!]
    \centering
    \begin{subfigure}[t]{0.48\linewidth}
        \centering
        \includegraphics[width=\linewidth]{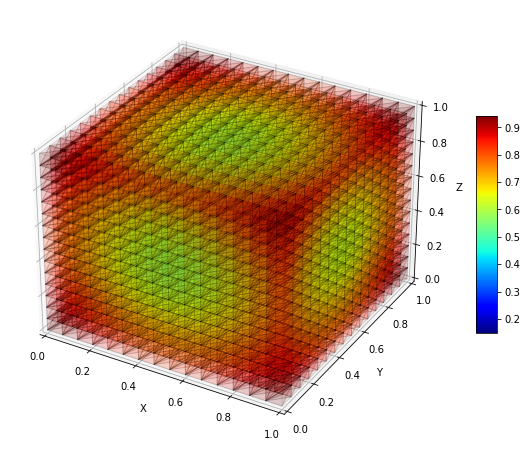}
        \caption{Input}
    \end{subfigure}
    \hfill
    \begin{subfigure}[t]{0.48\linewidth}
        \centering
        \includegraphics[width=\linewidth]{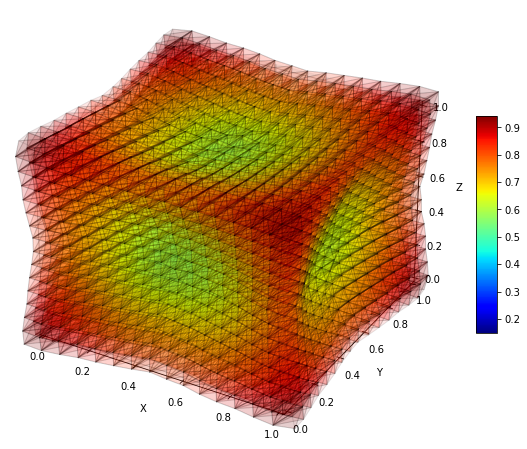}
        \caption{Output}
    \end{subfigure}
    \caption{The experimental result obtained by our proposed LDEM-3D method for the \emph{Spherical Shell Population Distribution} test case.}
    \label{fig:shell}
\end{figure}

\begin{figure}[t!]
    \centering
    \begin{subfigure}[t]{0.48\linewidth}
        \centering
        \includegraphics[width=\linewidth]{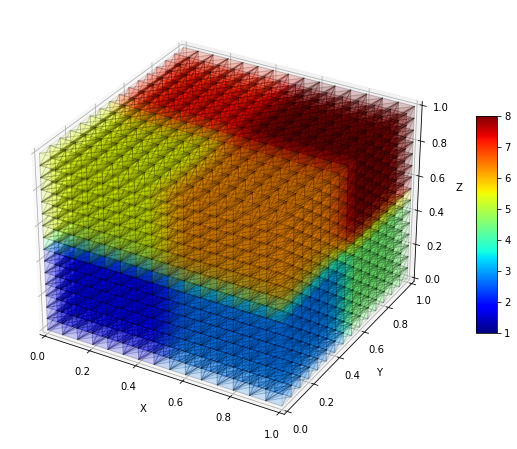}
        \caption{Input}
    \end{subfigure}
    \hfill
    \begin{subfigure}[t]{0.48\linewidth}
        \centering
        \includegraphics[width=\linewidth]{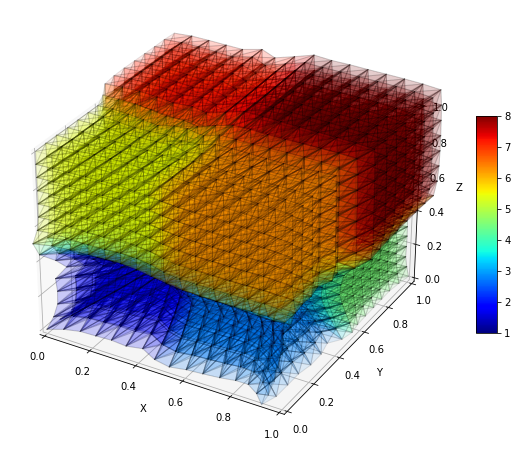}
        \caption{Output}
    \end{subfigure}
    \caption{The experimental result obtained by our proposed LDEM-3D method for the \emph{Smooth Blended Octants} test case.}
    \label{fig:smooth_blended_octants}
\end{figure}

More specifically, we consider four distinct test cases (Fig.~\ref{fig:basic_sinusoidal_3d}--\ref{fig:smooth_blended_octants}) to evaluate the model’s capability in handling diverse 3D population distributions. The \emph{Input} shows the initial 3D grid color-coded with the input population distribution, and the \emph{Output} shows the final 3D mapping result obtained by our method. 

In Fig.~\ref{fig:basic_sinusoidal_3d}, we first consider an example of \emph{3D Basic Sinusoidal Variation} generated using the sinusoidal functions as follows:
\begin{equation}
    \rho(\mathbf{c}) = 1.2 + \sin(2\pi c_x) \cdot \cos(2\pi c_y) \cdot \sin(2\pi c_z),
\end{equation}
where \(c_x, c_y, c_z\) denote the \(x\)-, \(y\)-, and \(z\)-coordinates of the centroid \(\mathbf{c}\). It can be observed that analogous to the 2D case, the proposed method can successfully create shape deformation with different regions enlarged or shrunk based on the input $\rho$.

Next, we consider the \emph{3D Complex Sinusoidal Variation} in Fig.~\ref{fig:complex_sinusoidal_3d}, in which a more intricate population pattern is defined using exponential and logarithmic transformations in 3D:
\begin{equation}
    \rho(\mathbf{c}) = 1.2 + \sin\big(\exp(c_x) \cdot 2\pi\big) \cdot \cos\big(\log(c_y + \epsilon) \cdot \pi\big) \cdot \sin(2\pi c_z),
\end{equation}
where \(\epsilon=10^{-5}\) is a small constant added to avoid singularity at \(c_y = 0\). This function produces sharper transitions and more irregular spatial structures. Again, an admissible mapping result with prominent shape deformation can be observed.

We then consider the \emph{Spherical Shell Population Distribution} example (Fig.~\ref{fig:shell}). Here, the distribution is defined by a central axis and a fixed radius \(R\):
\begin{equation}
    \rho(\mathbf{c}) = \exp\left(-\frac{(d(\mathbf{c}) - R)^2}{2 \cdot T^2}\right),
\end{equation}
where \(T\) controls the thickness of the shell. This simulates a spherical band of density concentrated around a 3D loop. It is easy to see that the central region of the 3D grid shrinks significantly in the mapping result, which matches the desired effect prescribed in $\rho$ very well.

Finally, we consider the \emph{Smooth Blended Octants} test case in Fig.~\ref{fig:smooth_blended_octants}. Specifically, to model smooth transitions across eight spatial regions (octants), we use a 3D extension of the smooth blending function:
\begin{equation}
    S(x; c, w) = \frac{1}{1 + \exp\left(-\frac{x - c}{w}\right)},
\end{equation}
where \(c = 0.5\) is the center of the transition and \(w = 0.02\) controls the sharpness of the blend. Then, given a point \(\mathbf{c} = (c_x, c_y, c_z)\), the population is defined as:
\begin{equation}
\begin{aligned}
\rho(\mathbf{c}) =\; & 1 \cdot \big(1 - S(c_x; 0.5, 0.02)\big) \cdot \big(1 - S(c_y; 0.5, 0.02)\big) \cdot \big(1 - S(c_z; 0.5, 0.02)\big) + \\
& 2 \cdot S(c_x; 0.5, 0.02) \cdot \big(1 - S(c_y; 0.5, 0.02)\big) \cdot \big(1 - S(c_z; 0.5, 0.02)\big) + \\
& 3 \cdot \big(1 - S(c_x; 0.5, 0.02)\big) \cdot S(c_y; 0.5, 0.02) \cdot \big(1 - S(c_z; 0.5, 0.02)\big) + \\
& 4 \cdot S(c_x; 0.5, 0.02) \cdot S(c_y; 0.5, 0.02) \cdot \big(1 - S(c_z; 0.5, 0.02)\big) + \\
& 5 \cdot \big(1 - S(c_x; 0.5, 0.02)\big) \cdot \big(1 - S(c_y; 0.5, 0.02)\big) \cdot S(c_z; 0.5, 0.02) + \\
& 6 \cdot S(c_x; 0.5, 0.02) \cdot \big(1 - S(c_y; 0.5, 0.02)\big) \cdot S(c_z; 0.5, 0.02) + \\
& 7 \cdot \big(1 - S(c_x; 0.5, 0.02)\big) \cdot S(c_y; 0.5, 0.02) \cdot S(c_z; 0.5, 0.02) + \\
& 8 \cdot S(c_x; 0.5, 0.02) \cdot S(c_y; 0.5, 0.02) \cdot S(c_z; 0.5, 0.02).
\end{aligned}
\end{equation}
This expression smoothly interpolates population values across all eight octants in the unit cube using only the coordinates \((c_x, c_y, c_z)\). From the mapping result, it is easy to see that the eight regions are enlarged or shrunk proportionally.

\begin{figure}[t!]
    \centering
    \includegraphics[width=0.8\textwidth]{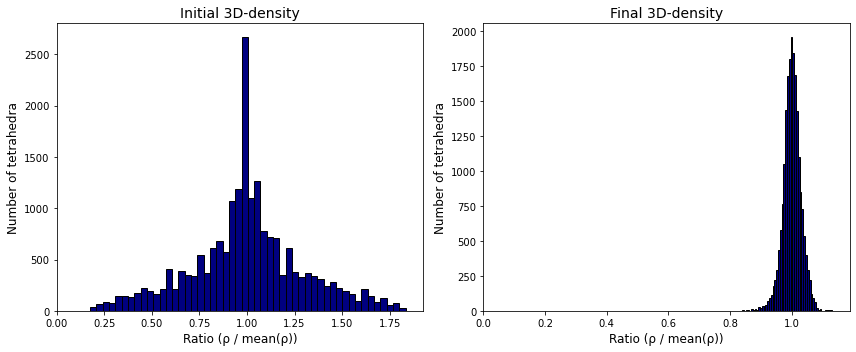}
    \caption{The histogram of the density distribution for the \emph{3D Basic Sinusoidal Variation} test case.}
    \label{fig:basic_distribution_3d}
\end{figure}

\begin{figure}[t!]
    \centering
    \includegraphics[width=0.8\textwidth]{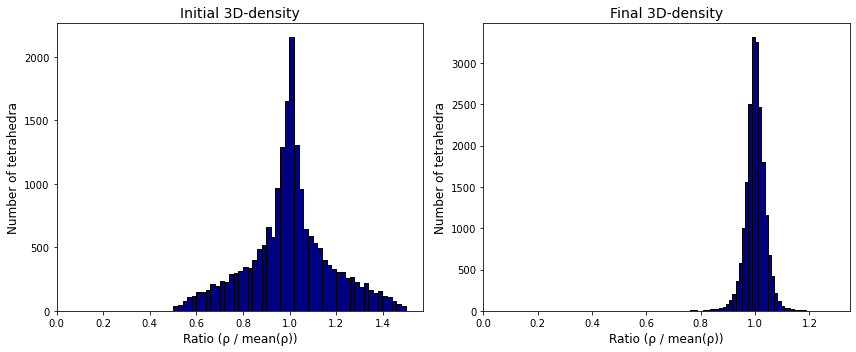}
    \caption{The histogram of the density distribution for the \emph{3D Complex Sinusoidal Variation} test case.}
    \label{fig:complex_distribution_3d}
\end{figure}

\begin{figure}[t!]
    \centering
    \includegraphics[width=0.8\textwidth]{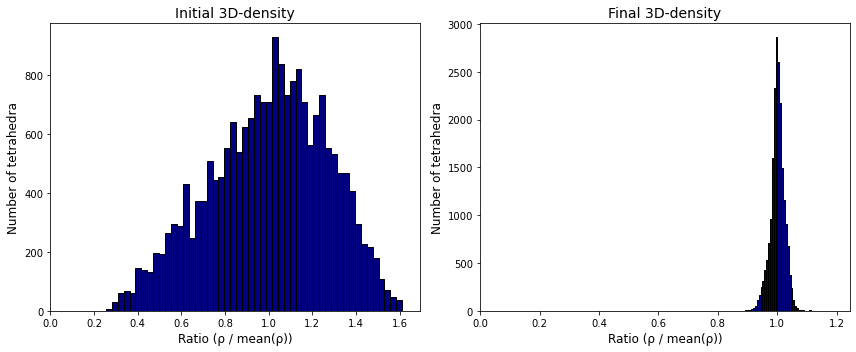}
    \caption{The histogram of the density distribution for the \emph{Spherical Shell Population Distribution} test case.}
    \label{fig:shell_distribution}
\end{figure}

\begin{figure}[t!]
    \centering
    \includegraphics[width=0.8\textwidth]{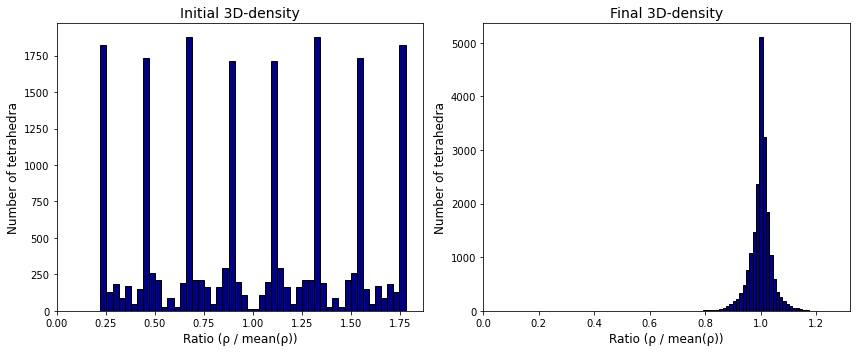}
    \caption{The histogram of the density distribution for the \emph{Smooth Blended Octants} test case.}
    \label{fig:blend_distribution_3d}
\end{figure}

For a more quantitative analysis, we also present the histograms of the initial density distribution ($\frac{\text{Population}}{\text{Initial volume}}$) and the final density distribution ($\frac{\text{Population}}{\text{Final volume}}$) for each test case (see Fig.~\ref{fig:basic_distribution_3d}--\ref{fig:blend_distribution_3d}). In all examples, it can be observed that the final density distribution is much more concentrated at 1 when compared with the initial density distribution. This shows that our LDEM-3D method can effectively achieve 3D density-equalizing maps with the desired shape deformation effects. 

\section{Conclusions}\label{sect:conclusion}

In this paper, we have developed a novel learning-based framework for constructing density-equalizing maps. By utilizing deep neural networks, we allow the model to generalize across arbitrary continuous population distributions, achieving robust performance without retraining for each new case. Our approach combines density equalization objectives with smoothness regularization, ensuring both fidelity and geometric plausibility of the resulting maps. Numerical experiments demonstrate the effectiveness of our method in producing accurate and visually coherent transformations for a wide range of prescribed population distributions. We have also applied our method for surface remeshing with different desired remeshing effects. Altogether, our method allows for scalable and robust computation of density-equalizing maps for practical applications.

As shown in the previous section, our proposed method can be easily extended from 2D to 3D for producing volumetric deformations based on prescribed population distributions. A natural next step is to further extend the proposed framework for more complex surface and volumetric domains to handle a wider class of shape mapping problems. We also plan to explore more of the theoretical properties of the learned mappings in our future work to better understand their limitations.

\bibliographystyle{ieeetr}
\bibliography{reference.bib}

\end{document}